\documentclass[fleqn,usenatbib,useAMS]{mnras}

\usepackage{graphicx}
\usepackage{amsmath}
\usepackage{amssymb}    
\usepackage{multicol}      
\usepackage{bm}
\usepackage{float}
\usepackage{chngcntr}
\usepackage{soul}
\usepackage{pdflscape}    
\usepackage{caption}
\usepackage{subcaption}
\usepackage{enumitem}
\usepackage{booktabs}
\usepackage{rotating}
\usepackage{adjustbox}
\usepackage[toc,page]{appendix}
\usepackage{pdflscape}
\usepackage{subcaption}
\usepackage{natbib}
\usepackage{xcolor}
\usepackage{rotating}
\defcitealias{pubA}{Publication~A}
\defcitealias{pubB}{Publication~B}
\newcommand{\oiii}[1]{\text{[O\,III]$\lambda$5007}\if\relax\detokenize{#1}\relax\else\xspace\fi#1}
\newcommand{\nii}[1]{\text{[N\,II]$\lambda$6583}\if\relax\detokenize{#1}\relax\else\xspace\fi#1}
\newcommand{\ha}[1]{\text{H$\alpha$\if\relax\detokenize{#1}\relax\else\xspace\fi#1}}
\newcommand{\hb}[1]{\text{H$\beta$\if\relax\detokenize{#1}\relax\else\xspace\fi#1}}

\mathchardef\mhyphen="2D

\title[Extreme cloud collisions in nearby barred galaxies]{Extreme cloud collisions in nearby barred galaxies}

\author[Kolcu et al.]{Tutku Kolcu$^{1,2}$\thanks{Contact e-mail: \href{mailto:T.Kolcu@2020.ljmu.ac.uk}{Tutku.Kolcu@nottingham.ac.uk}}, Mattia C.~Sormani$^{3}$, Witold Maciejewski$^{1}$, Sophia K. Stuber$^4$, Eva Schinnerer$^{4}$, \newauthor{Francesca Fragkoudi$^{5}$}, Ashley T. Barnes$^{6}$, Frank Bigiel$^{7}$, Mélanie Chevance$^{8, 9}$, Dario Colombo$^{7}$, \newauthor{Éric Emsellem$^{6,10}$}, Simon C. O. Glover$^{8}$, Jonathan D. Henshaw$^{4, 1}$, Ralf S. Klessen$^{8, 11, 12, 13}$, \newauthor{Sharon E. Meidt$^{14}$}, Justus Neumann$^{4}$, Francesca Pinna$^{15,16}$,  Miguel Querejeta$^{17}$, \newauthor{Thomas G. Williams$^{18}$}  \\
$^{1}$ School of Physics and Astronomy, University of Nottingham, University Park, Nottingham NG7 2RD, UK \\
$^{2}$ Astrophysics Research Institute, Liverpool John Moores University, IC2 Liverpool Science Park, 146 Brownlow Hill, L3 5RF, UK \\
$^{3}$ Como Lake centre for AstroPhysics (CLAP), DiSAT, Universit{\`a} dell’Insubria, via Valleggio 11, 22100 Como, Italy \\
$^{4}$ Max-Planck-Institut für Astronomie, Königstuhl 17, 69117 Heidelberg, Germany \\
$^{5}$ Institute for Computational Cosmology,  Department of Physics, Durham University, DH1 3LE, United Kingdom \\
$^{6}$ European Southern Observatory, Karl-Schwarzschild 2, 85748
Garching bei Muenchen, Germany \\ 
$^{7}$ Argelander-Institut für Astronomie, Universität Bonn, Auf dem
Hügel 71, 53121 Bonn, Germany \\
$^{8}$ Institut für Theoretische Astrophysik, Zentrum für Astronomie der
Universität Heidelberg, Albert-Ueberle-Strasse 2, 69120 \\ Heidelberg, Germany \\
$^{9}$ Cosmic Origins Of Life (COOL) Research DAO, coolresearch.io \\
$^{10}$ Univ Lyon, Univ Lyon 1, ENS de Lyon, CNRS, Centre de
Recherche Astrophysique de Lyon UMR5574, 69230 Saint-Genis-Laval, France \\
$^{11}$ Universität Heidelberg, Interdisziplinäres Zentrum für Wissenschaftliches Rechnen, Im Neuenheimer Feld 225, 69120, \\Heidelberg, Germany \\
$^{12}$ Harvard-Smithsonian Center for Astrophysics, 60 Garden Street, Cambridge, MA 02138, U.S.A \\
$^{13}$ Elizabeth S. and Richard M. Cashin Fellow at Radcliffe Institute for Advanced Studies at Harvard University, 10 Garden Street, Cambridge, MA 02138, USA \\
$^{14}$ Sterrenkundig Observatorium, Universiteit Gent, Krijgslaan 281 S9, B-9000 Gent, Belgium \\
$^{15}$ Instituto de Astrofísica de Canarias, C/ Vía Láctea s/n, E-38205, La Laguna, Spain \\
$^{16}$ Departamento de Astrofísica, Universidad de La Laguna, Av. del Astrofísico Francisco Sánchez s/n, E-38206, La Laguna, Spain \\
$^{17}$ Observatorio Astronómico Nacional (IGN), C/Alfonso XII 3, Madrid 28014, Spain \\
$^{18}$ Sub-department of Astrophysics, Department of Physics, University of Oxford, Keble Road, Oxford OX1 3RH, UK
}

\date{Accepted XXX. Received YYY; in original form ZZZ}

\begin{document}

\maketitle

\begin{abstract}
The inner regions of the Milky Way are known to contain an enigmatic population of prominent molecular clouds characterised by extremely broad lines. The physical origin of these ``extended velocity features'' (EVFs) is still debated, although a connection with the ``dust lanes'' of the Galactic bar has been hypothesised. In this paper, we search for analogous features in the dust lanes of nearby barred galaxies using the PHANGS-ALMA CO(2-1) survey. We aim to confirm existence of EVFs in other galaxies and to take advantage of the external perspective to gain insight into their origin. We study a sample of 29 barred galaxies and find that 34\% contain one or more EVFs, while the remaining lack obvious signs of EVFs. Upon analysing the physical properties of the EVFs, we find they possess large virial parameters, ranging from few hundreds to several thousand, indicating that they are strongly out-of-equilibrium. The most likely explanation for their origin is extreme cloud-cloud collisions with relative velocities in excess of 100km/s in highly non-circular flow driven by the bar. This interpretation is consistent with previous high-resolution observations in Milky Way. Further corroboration of this interpretation comes from the inspection of high-sensitivity infrared observations from the PHANGS-JWST Treasury Survey that reveals streams of gas that appear to be hitting the dust lanes at locations where EVFs are found. We argue that EVFs are the clearest examples of cloud-cloud collisions available in literature and represent a unique opportunity to study cloud collisions and their impact on star formation.
\end{abstract}

\section{Introduction}
\label{sec:intro}

\begin{figure*}
    \centering
    \includegraphics[width=1\textwidth]{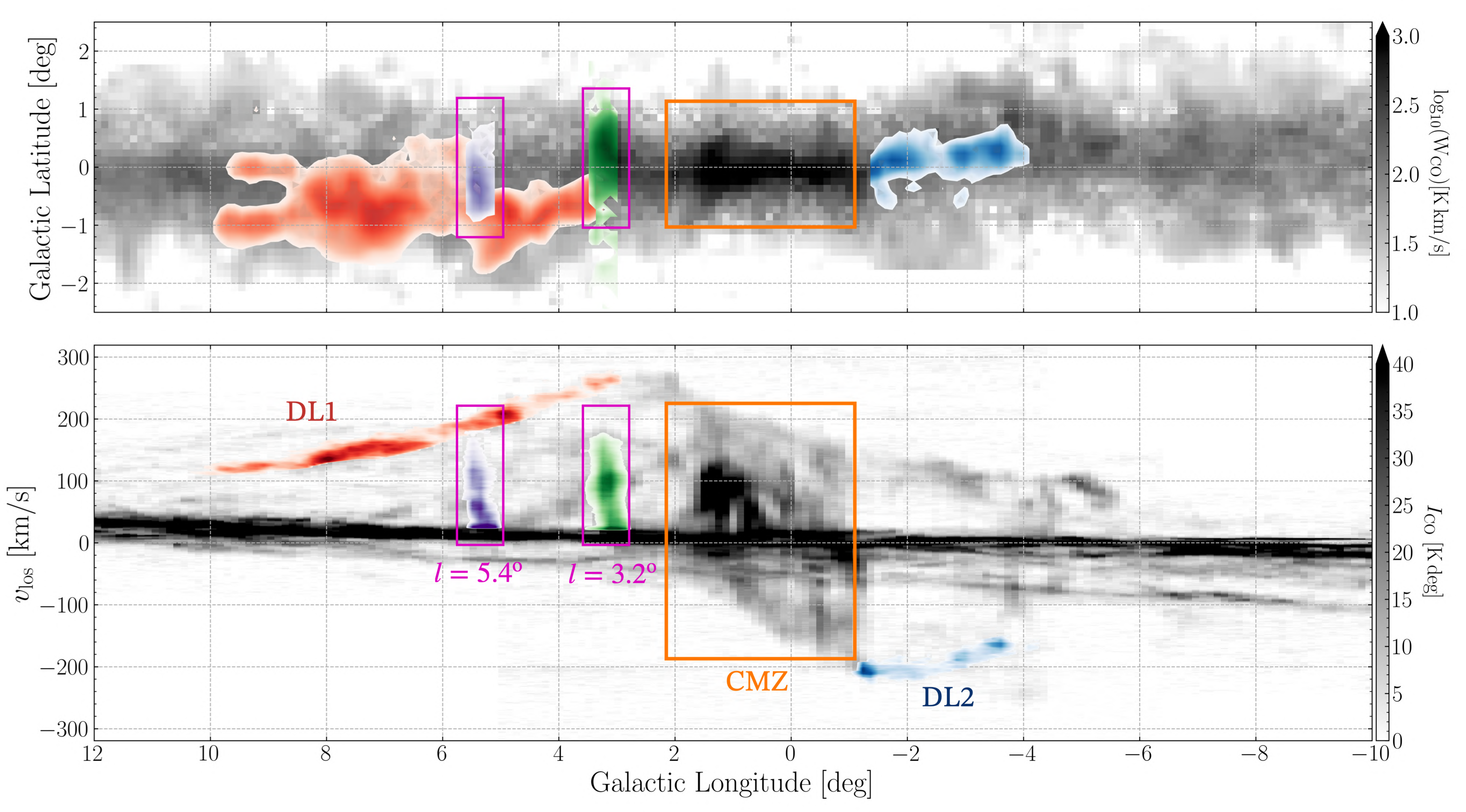}
    \caption{CO(1-0) emission maps of the central regions of the Milky Way. The top and bottom panels display the longitude-latitude $(l,b)$ and longitude-velocity $(l,v_{\rm los})$ CO(1-0) emission maps respectively from the survey by \citet{Bitran_1997}. The two most prominent EVFs of the Galaxy, located at $l=5.4^\circ$ and $l=3.2^\circ$, are marked with magenta boxes and are coloured in indigo and green respectively. Red and blue are the emission from the bar dust lanes on the redshifted (DL1) and blueshifted (DL2) sides of the Galaxy. The orange box marks the Central Molecular Zone (CMZ). The Galactic centre is located at $(l,b) = (0,0)$.}
    \label{fig:MilkyWay}
\end{figure*}

It has been known for several decades that the innermost few kpc of the Milky Way (MW) contain an enigmatic population of prominent molecular clouds characterised by extremely broad lines \citep[spanning velocity ranges in excess of $100\mhyphen 200 $\kms, e.g.][]{Bitran_1997,Kumar1997,Liszt_2006}. These peculiar objects are found only in the vicinity of the Galactic centre (Galactic longitudes $|l|<6^\circ$), in the region where the gravitational potential is dominated by the Galactic bar, and can be immediately identified in the longitude-velocity diagrams as near-vertical features (see Fig.~\ref{fig:MilkyWay} below or Fig.~2 in \citealt{Henshaw2023}). Following \cite{Sormani_2019} and \cite{Henshaw2023}, we will refer to these objects as Extended Velocity Features (EVFs).  

A number of diverse interpretations have been put forward for the origin of EVFs. \citet{Stark_1986}, \citet{Boyce_1989}, \citet{Lee_1999} and \citet{Baba_2010} suggested that they are streams of clouds falling towards the Galactic centre, although as \cite{Stark_1986} acknowledged, this explanation requires the clouds to be near-perfectly aligned along the line-of-sight (the so-called ``fingers of God" effect). \citet{Fukui_2006} (see also \citealt{Fujishita_2009,Torii_2010,Suzuki2015,Enokiya2021}) argued that they are the footprints of giant magnetic loops that are lifted above the Galactic plane by magnetic buoyancy through the \citet{Parker_1966} instability.

Several authors have proposed that the EVFs are related to gas flow in the non-axisymmetric gravitational potential generated by the Galactic bar, with the interpretation being modified and refined over the years. \cite{Kumar1997} suggested that the velocity extent of the EVFs is simply turbulent velocity dispersions caused by cloud-cloud collisions in the bar potential. \citet{Fux_1999} and \cite{Liszt_2006} modified this interpretation and suggested that EVFs are lumps of gas that are traversing the ``dust lanes'' of the Milky Way bar. According to these authors, the velocity extent is not simply a velocity dispersion but reflects a steep velocity gradient due to the rapid differential acceleration of the cloud as it crosses the dust lanes, which from a hydrodynamical point of view are large-scale shocks in the gas flow and therefore locations where the velocity changes abruptly \citep{Athanassoula_1992,Piner_1995,Fragkoudi_16}. \cite{Sormani_2019} further refined this interpretation by showing that EVF-like features are naturally produced in simulations of gas flow in barred potentials. EVFs arise when a stream of gas travelling towards the Galactic centre along one of the dust lanes collides with 1) a stream of gas coming from the other dust lane on the opposite side of the Galaxy that has ‘overshot’ the central molecular zone (CMZ) gas ring; or 2) a stream of gas orbiting in the CMZ ring, which has orbital velocities substantially lower than the material infalling along the dust lane. Recent high-resolution zoom-in observations of the MW's EVFs are consistent with this picture and show velocity bridges that can be interpreted as a direct interaction between two colliding clouds/streams with very different bulk velocities \citep{Busch2022,Gramze_2023}.

Despite these efforts, it has remained difficult to conclusively confirm or disprove the above bar-related interpretation due to our embedded perspective within the Milky Way disc. In this paper, we therefore turn our attention to external galaxies and search for EVF analogues in nearby barred galaxies. We aim to confirm the existence of EVFs in external galaxies and use the vantage provided by our external point of view to gain insight into their physical origin. Since the extent of the largest EVFs in the MW is approximately $0.7^\circ \sim100$\,pc at the distance of the Galactic centre, this program requires observations of molecular clouds in nearby galaxies at resolutions on the order of $\sim100$ pc or better. These have only recently become available thanks to data obtained by the \textit{Physics at High Angular resolution in Nearby GalaxieS (PHANGS)} collaboration, in particular the PHANGS-ALMA survey \citep{Leroy_21b} that has mapped ${}^{12}$CO $J=2\to 1$ line emission in 90 nearby (distance $\leq$ 20\,Mpc) massive star-forming galaxies at a spatial resolution of 50-150 pc, and the PHANGS-JWST Treasury Survey \citep{Lee_23} that has obtained infrared imaging of many of the same galaxies at a resolution of 5--50 pc.

The paper is structured as follows. In Sect.~\ref{sec:obs_data}, we describe the PHANGS--ALMA and PHANGS--JWST data. In Sect.~\ref{sec:diagnostics_EVF_properties}, we describe the construction of the diagnostic maps used in our study to identify EVFs. In Sect.~\ref{sec:results} we present our results. We provide conclusions in Sect.~\ref{sec:conclusion}.

\section{Observational data}
\label{sec:obs_data}

\subsection{PHANGS-ALMA CO(2--1) Survey}
\label{sec:ALMA}

We use the $^{12}$CO(2--1) datacubes from the PHANGS–ALMA survey \citep{Leroy_21b}. The survey mapped the \mbox{CO(2--1)}  emission of 90 nearby (distance $<$ 20\,Mpc) star-forming galaxies in the southern sky. 
Observations of 58 galaxies within the PHANGS-ALMA sample were taken as a part of Cycle 5 ALMA Large Program. Observations of the remaining galaxies were taken as a part of several smaller programmes. The position-position-velocity (PPV) datacube has a velocity resolution of 2.5\,km\,s$^{-1}$ and an angular resolution of $\sim$1\arcsec, which corresponds to a typical spatial resolution of $\sim$\,50--150\,pc depending on the distance to the galaxy. For more details on the PHANGS-ALMA Survey data, see \citet{Leroy_21b} and for an in-depth discussion of the data reduction, pipeline and imaging see \citet{Leroy_21a}. 

The PHANGS-ALMA Survey is a ``cloud-scale" spectroscopic survey in the sense that the observations have the resolution, point-source sensitivity, and beam diameter sufficient to resolve and study individual giant molecular clouds (GMCs). These GMCs typically have radii around 30 to 60\,pc and characteristic molecular masses on the order of $\rm 10^5M_\odot$ \citep{Bolatto_08, Freeman_17, Rosolowsky_21,Schinnerer_2024}. The provided data offers sufficient spatial and velocity resolution for our study aimed at identifying features that have typical sizes comparable to the most prominent EVFs of the MW ($\sim 100$ pc, see introduction).

As discussed in \citet{Stuber_2023} (see their Sect. 5.1 and Appendix C), the molecular gas response to the underlying gravitational potential can lead to morphological features that differ from those observed in stellar light. This can result in significant discrepancies between classifications based on photometric surveys, such as the Spitzer Survey of Stellar Structure in Galaxies (S$^4$G; \citealp{Buta_2015}), and those based on molecular gas morphology. For example, some galaxies classified as strongly barred in the literature appear unbarred in CO(2–1), potentially indicating gas-depleted bars where further inflow has ceased, or cases where star-forming clumps mimic elongated features in stellar light but lack coherent gas structures in CO. Conversely, certain galaxies (e.g., NGC 4941) show clear bar-like features in CO that are absent or ambiguous in optical and infrared images, suggesting that molecular gas can reveal structures that are not easily detectable in stellar light.

Therefore, to ensure that our selected galaxies exhibit a bar-like feature in molecular gas, in this work we use a sample of 29 galaxies from the PHANGS-ALMA dataset, listed in Table 1, all classified as barred, based on their CO(2–1) morphology, by \citet{Stuber_2023} (class B and C there). This selection lessens the potential biases introduced by stellar light-based classifications and ensures a consistent approach in studying bar-driven processes in molecular gas. For each galaxy, we use the ALMA CO(2--1) datacubes to identify EVFs.

\begin{table*}
\centering
\caption{Sample of barred galaxies considered in this work. Columns are: (1) Galaxy ID; (2) Distances obtained from \citet{Anand_2021} (3) Physical resolution in native resolution cube provided in \citet{Leroy_2021}; (4) inclination -- ``..." indicates that the uncertainties are not provided and (5) kinematic position angle of LON from \citet{Lang_2020} when available, otherwise, from \citet{Salo_2015} (entries from the latter are marked with an asterisk, $*$); (6) position angle of the bar provided in the S$^4$G Survey \citet{Buta_2015, Herrera-Endoqui_2015} or \citet{Salo_2015} ; (7) Bar class assigned by \citet{Stuber_2023} based on CO morphology; (8) Photometric bar classification in literature obtained from \citet{Buta_2015} (NIR) if available or from \citet{deVaucouleurs_1991} (optical); (9) Presence of EVFs identified in this work, based on PP and PV diagrams, as discussed in Sect.\ref{sec:results}
.`Y' indicates that EVF signatures are found in the galaxy, while `N' indicates no clear signatures of EVFs are detected.}
\begin{tabular}{|c|c|c|c|c|c|c|c|c|}
\hline
{Galaxy ID} & Distance &  Resolution & $i$ & PA of LON & PA of bar & Bar Type & Bar Type & EVF \\ & (Mpc) & (pc) & (deg) & (deg) & (deg) & (CO morph) & (Photometric) & (Y,N) \\
 (1) &(2) & (3) & (4) & (5) & (6) & (7)  & (8) & (9) \\
\hline
NGC\,0253 & 3.70& 150.2  & 75.0$\pm$... & 52$^*$ & 66 & B & SAB & N \\
NGC\,0685 & 19.94& 163.0 & 23.0$\pm$43.4 & 101 & 55 & B  & SB & N \\
NGC\,1087 & 15.85& 123.1 & 42.9$\pm$ 3.9 & 179 & 128 & B & SB & N \\
NGC\,1097 & 13.58& 111.7 & 48.6$\pm$6.0 & 122 & 141 & C  & SB & \textbf{Y} \\
NGC\,1300 & 18.99& 113.1 & 31.8$\pm$6.0 & 98 & 99 & C & SB & \textbf{Y} \\
NGC\,1317 & 19.11& 147.1& 23.2$\pm$7.7 & 42 & 150 & B  & SB & N \\
NGC\,1365 & 19.57& 130.8& 55.4$\pm$6.0 & 21 & 86 & C & SB & N \\
NGC\,1433 & 18.63& 99.1& 28.6$\pm$6.0 & 20 & 95 & C & SB & N \\
NGC\,1512 & 18.83& 94.5& 42.5$\pm$6.0 & 82 & 42 & C & SB & \textbf{Y} \\
NGC\,1566 & 17.69& 107.6& 29.6$\pm$10.7 & 35 & 177 & C & SAB & N \\
NGC\,1637 & 11.7& 78.9& 31.0$\pm$...  & 21$^*$& 85 & B & SAB & N \\
NGC\,1672 & 19.4& 181.7& 42.6$\pm$6.0 & 134 & 96 & C  & SA\underline{B} & N \\
NGC\,2566 & 23.44& 145.3& 48.5$\pm$6.0 & 132 & 66 & C & SB & \textbf{Y} \\
NGC\,2903 & 10.0& 70.5& 66.8$\pm$3.1 & 24 & 28 & C& SB & \textbf{Y} \\
NGC\,3059 & 20.23& 119.9& 29.4$\pm$11.0 & 165 & 40 & B & SB & N \\
NGC\,3351 & 9.96& 70.7& 45.1$\pm$6.0 & 13 & 112 & C  & SB & N \\
NGC\,3507 & 23.55& 155.2& 21.7$\pm$11.3 & 56 & 112 & C & SA\underline{B} & N \\
NGC\,3627 & 11.32& 89.2& 57.3$\pm$1.0 & 173 & 160 & C & SB & \textbf{Y} \\
NGC\,4303 & 16.99& 149.3& 23.5$\pm$9.2 & 132 & 178 & C & SAB & \textbf{Y} \\
NGC\,4321 & 15.21& 122.9& 38.5$\pm$2.4 & 156 & 115 & C  & SAB & N \\
NGC\,4535 & 15.77& 119.1& 44.7$\pm$10.8 & 180 & 42 & C & SAB & \textbf{Y} \\
NGC\,4548 & 16.22& 132.7& 38.3$\pm$6.0 & 138 & 61 & C & SB & \textbf{Y} \\
NGC\,4579 & 21.00& 182.7& 40.2$\pm$5.6 & 91 & 53 & C  & SB & N \\
NGC\,4654 & 21.98& 182.8 & 55.6$\pm$5.9 & 123 & 121 & B  & SB & N \\
NGC\,4941 & 15.00& 115.3& 53.4$\pm$1.1 & 22 & 16 & C  & SA & N \\
NGC\,5236 & 4.89& 50.7& 24.0$\pm$... & 45$^*$ & 50 & C  & SAB & N \\
NGC\,5643 & 12.68& 79.9& 29.9$\pm$6.0 & 139 & 84 & C & SAB & \textbf{Y} \\
NGC\,6300 & 11.58& 60.4& 49.6$\pm$5.9 & 105 & 71 & C & SB & N \\
NGC\,7496 & 18.72& 152.0& 35.9$\pm$0.9 & 14 & 147 & C & SB & N \\
\hline
\end{tabular}

\label{tab:PHANGS_galaxysample_barred}
\end{table*}

\subsection{PHANGS-JWST Treasury Survey}
\label{sec:jwst_obs}

We complement the PHANGS-ALMA CO(2--1) observations with data from the PHANGS-JWST Treasury Survey \citep{Lee_23}, which provides observations from the Mid-Infrared Instrument (MIRI) of JWST for 19 galaxies in Cycle 1 and 55 galaxies in Cycle 2. Among these, observations of 18 out of the 29 galaxies in our sample are included within their public data release. Further information on the Cycle 1 observations of the Treasury Survey can be found in \citet{Lee_23}, and more details on the PHANGS-JWST data release, data reduction, and processing can be found in \citet{Williams_2024}. 

The mid-IR emission captured by MIRI bands provides key insights into the interstellar medium (ISM) by acting as a tracer of \emph{both} gas column densities and star forming regions \citep{Leroy_2023}. In this work, we use the 770W filter, which is sensitive to 7.7\,$\rm \mu m$ emission from polycyclic aromatic hydrocarbons (PAHs) \citep{Smith_2007, Li_2020}. PAHs produce distinctive emission features when they are heated by ultraviolet (UV) photons. Near star-forming regions, the UV radiation field is usually dominated by radiation from young stars, but away outside star forming regions it also includes a contribution from the interstellar radiation field. As a result, the 7.7\,$\rm \mu m$ emission has been shown to correlate well with both H$\alpha$ emission (which traces star forming regions) and CO gas (which traces molecular gas), so that this emission is a superposition of a CO-tracing component and H$\alpha$-tracing component in roughly equal proportions \citep{Leroy_2023}. Crucially for this work, the 7.7\,$\rm \mu m$ emission has a strong correlation to the CO (2-1) emission away from star forming regions, allowing us to detect CO streams that go undetected in the CO-ALMA data because they fall below the sensitivity limit of the latter.

\section{EVF identification and property estimation}\label{}
\label{sec:diagnostics_EVF_properties}
\subsection{Diagnostic maps and EVF identification} \label{sec:diagnostic}

To search for EVFs in each of the 29 galaxies of our sample, we construct one position position (PP) and two position-velocity (PV) maps from its PHANGS-ALMA \mbox{CO(2--1)} data cube by integrating the cube over velocities and over two spatial dimensions, respectively. Before the integration, the cube is rotated so that the positive half of the line of nodes is along the negative x-axis. In this work, PAs of the LONs for majority of our sample galaxies are taken from \citet{Lang_2020} when available, otherwise taken from \citet{Salo_2015} (see Table \ref{tab:PHANGS_galaxysample_barred}). Then we convert angular (arcseconds) into physical (kpc) scales assuming the distances from \citet{Anand_2021}, so that $x$ and $y\cos(i)$ are the spatial coordinates in kpc in the sky plane (corresponding to $x$ and $y$ coordinates in the disc plane). 

To identify EVFs, in analogy with the MW's EVFs we look for features on the dust lanes that have an unusually large velocity spread but are relatively compact in size. Thus we first visually identify the dust lanes in the \mbox{CO(2--1)} PP maps. These typically appear unambiguously as two nearly straight or curved features, located on either side of the bar, extending from the nucleus to the bar ends (see Sect.~3.2.2 of \citealt{Stuber_2023} for further details). To better visualise each dust lane we draw a contour enclosing it in the PP map, and we colour in red or blue all data in the data cube with spatial coordinates falling within this contour. For this purpose we use the visualisation software {\texttt{glue}\footnote{The \texttt{glue} webpage can be accessed via \url{https://glueviz.org}}  \citep{Beaumont_2015, Robitaille_2017}. 
While this procedure is somewhat subjective, the results are reasonable, as the dust lanes are typically coherent and continuous structures in PP space. The top and middle panels of Fig.~\ref{fig:NGC1300} show the results of this procedure for NGC\,1300. In the following, we will refer to these diagrams as ``diagnostic maps". 

We then visually select the EVFs within the extracted dust lanes by looking in the PV maps for features that are abnormally broad-lined compared to the rest of the material on the dust lane, and that have physical sizes in the PP maps smaller or comparable to the width of the dust lane. Each selected EVF is outlined by rectangular boxes in PP and PV maps in Fig.~\ref{fig:NGC1300} and Figs.~\ref{fig:NGC1097}--\ref{fig:NGC5643}, and if more than one EVF is present in a galaxy, the EVFs are named F1, F2, etc., with the names plotted in the figures and listed in Table \ref{tab:EVF_candidates}.

The EVFs identified in this way typically look like ``blobs'' with an approximately circular shape in PP space, likely due to the resolution limitations. In several galaxies, as discussed in Sect.~\ref{sec:results}, these blobs are resolved into smaller clumps in 7.7\,$\rm \mu m$ emission (see Fig.~\ref{fig:JWST-Zoom-NGC1300} for NGC\,1300). These clumps might host incipient star formation activity, potentially triggered by molecular cloud collisions.

Only the PHANGS-ALMA CO(2-1) data is used to identify EVFs. The mid-infrared PHANGS-JWST observations are not utilized for EVF identification, but they are used to further inspect the vicinity of a CO-identified EVF and trace gas that falls below the sensitivity threshold of the CO data. As we shall see, this often reveals streams of gas connecting to the EVF that are invisible in the CO data. We focus only on the 7.7\,$\rm \mu$m wavelength from MIRI/JWST as these are available for all galaxies in our sample that exhibit EVFs and have been shown to trace well the gas column density \citep{Leroy_2023}. The bottom panel of Fig.~\ref{fig:NGC1300} shows an example JWST mid-infrared map for NGC\,1300.

The diagnostic maps and 7.7\,$\rm \mu$m images of the remaining galaxies in our sample are presented in Appendix \ref{sec:app_diagnostics_sample}.

\begin{figure*}
    \hspace{-0.2cm}
    \includegraphics[width=1\textwidth]{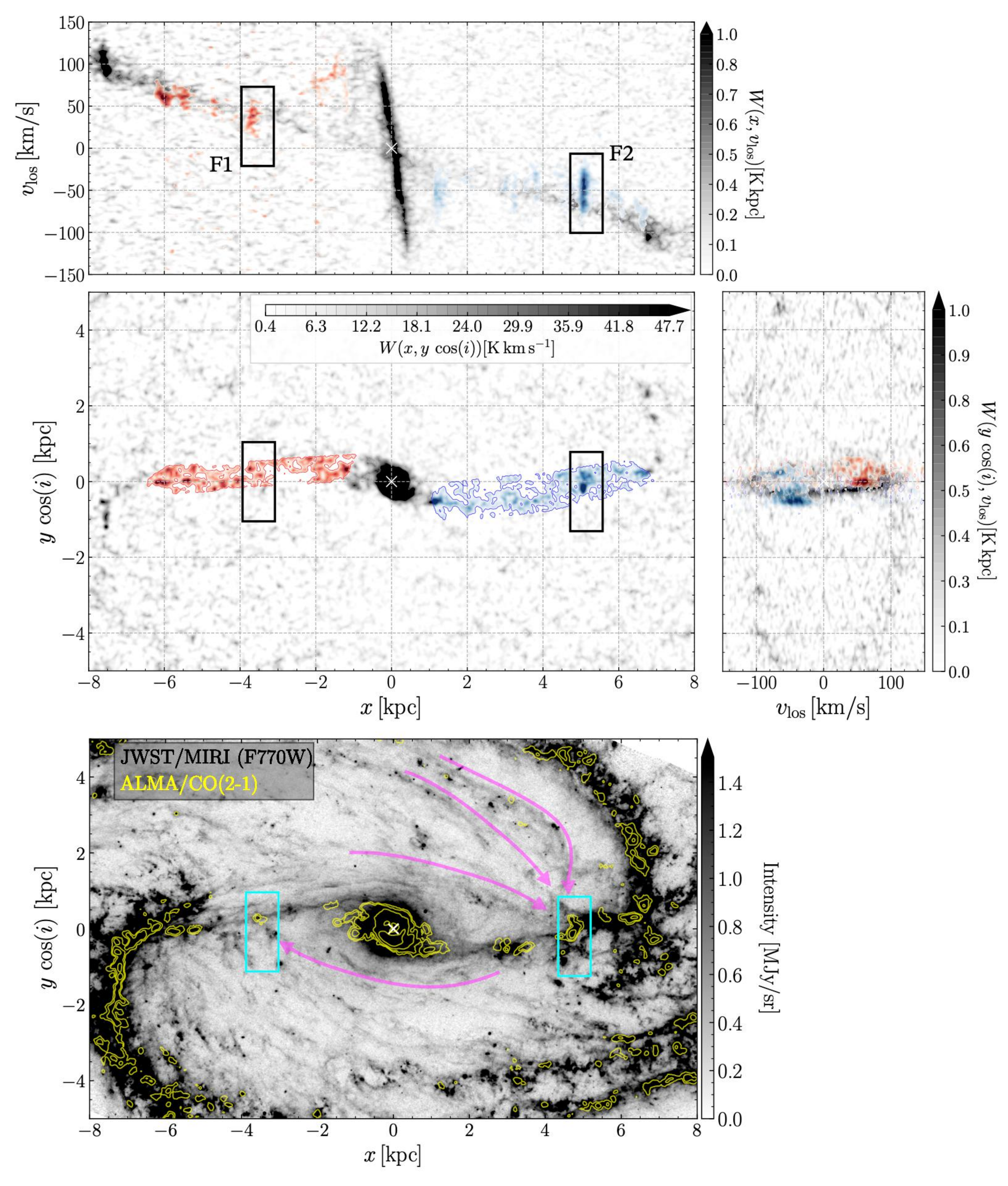}
    
    \caption{The diagnostic maps and 7.7$\mu$m emission image of NGC\,1300. \textit{Top:} Position-position (PP) and Position-velocity (PV) diagrams. The galaxy is rotated in the plane of the sky such that the PA of the LON aligns with the \textit{x}-axis. Red and blue colours highlight the dust lanes. 
    \textit{Bottom:} 7.7$\mu$m emission in gray-scale, with CO(2--1) intensity as in PP diagram in yellow contours. Magenta lines mark the streams that appear to extend over large radii and reach to the regions where EVFs are seen. In all panels, the rectangular boxes mark the regions of EVFs.}
    \label{fig:NGC1300}
\end{figure*}

\begin{figure*}
    \includegraphics[width=0.7\textwidth]{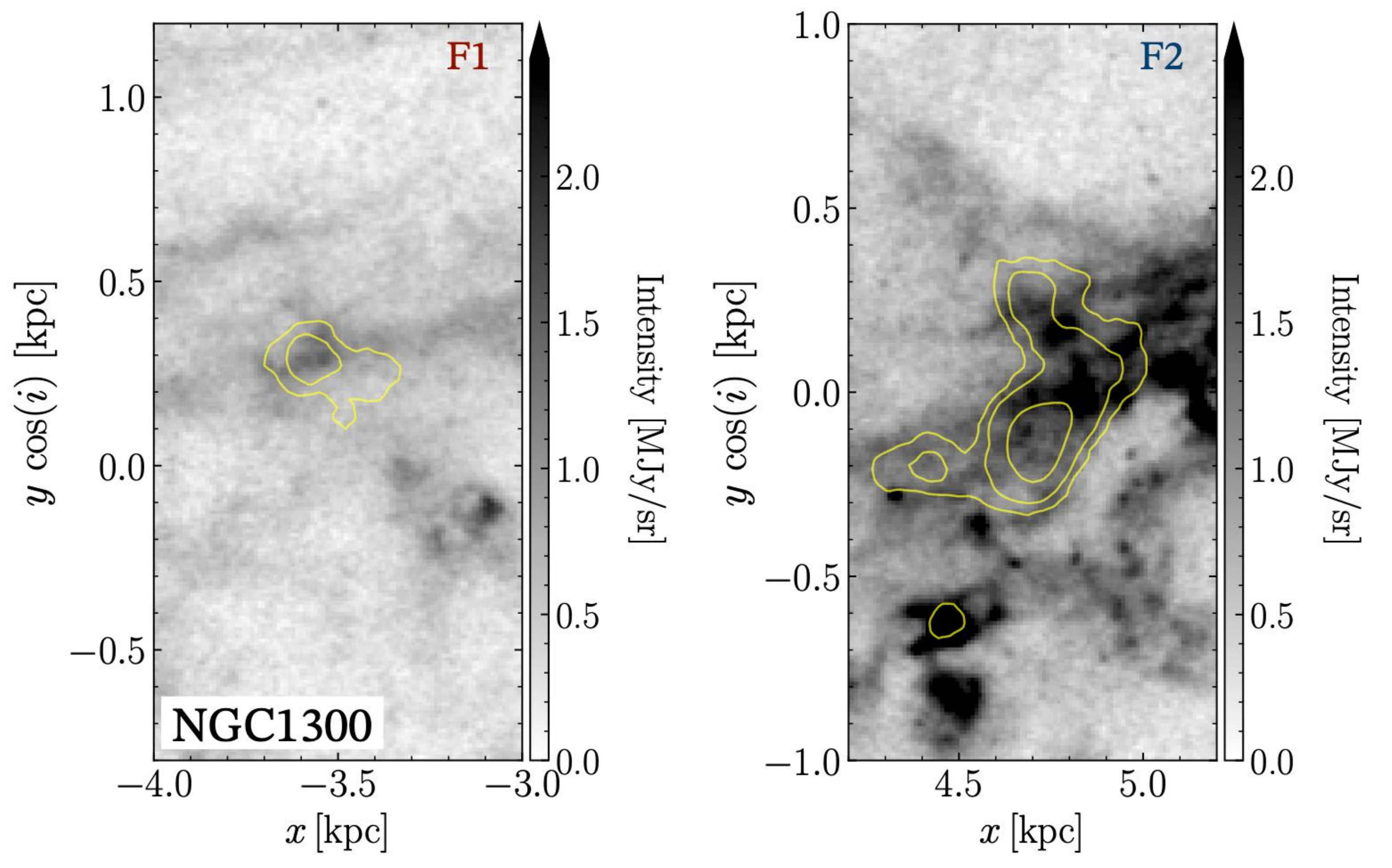}
\caption{Close-up views of the EVF regions, marked with rectangular boxes in the bottom panel of Fig.~\ref{fig:NGC1300}. Both panels show the 7.7\,$\rm \mu m$ emission in gray-scale overlaid with CO(2–1) emission in yellow contours.}
\label{fig:JWST-Zoom-NGC1300}
\end{figure*}

\begin{figure*}
    \centering
    \includegraphics[width=1\textwidth]{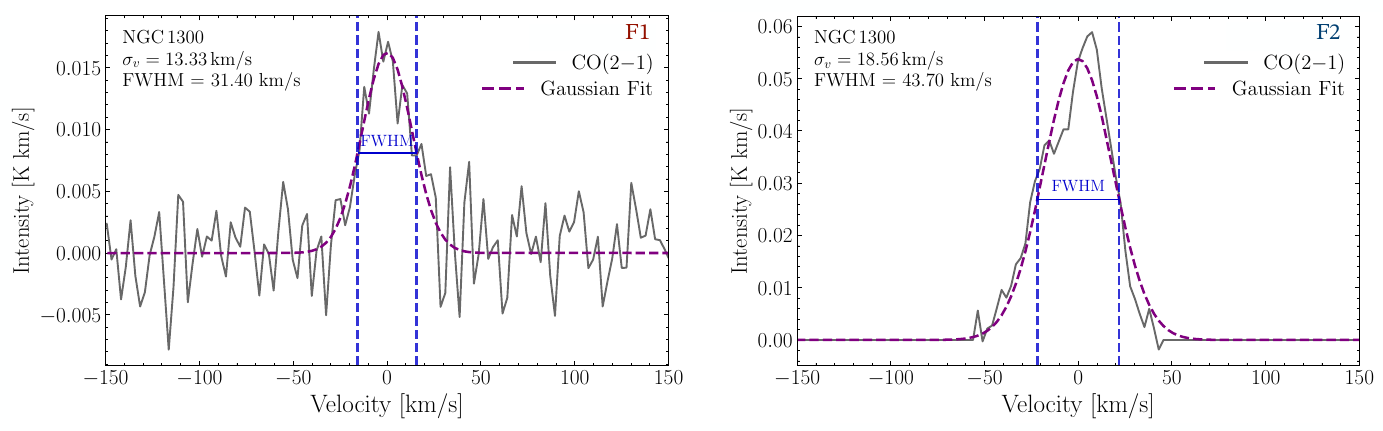}
    \caption{The CO(2--1) spectra within the regions enclosed within rectangular boxes in Fig.~\ref{fig:NGC1300} where EVFs are seen, for F1 on left and for F2 on right. Gray solid lines shows the total intensity, the purple dashed lines are the Gaussian fit to the spectrum and blue dashed lines mark the FWHM of the fits.}
    \label{fig:NGC1300_spectrum}
\end{figure*}

\subsection{EVF properties estimation}
\label{sec:properties_estimation}

To estimate the physical size of the EVFs, we first visually define a circular region on the dust lane, in PP space, that encompasses the entire EVF. We make the radius of the circle big enough to contain the whole EVF and small enough that it excludes (as much as possible) other nearby unrelated features. We also ensure that the circular region remains entirely within the dust lane. This region is then extracted and isolated for further analysis.

We estimate the EVF radius as follows. We centre the extracted region on the brightest point as an initial guess for the EVF centre. We then fit a 2D Gaussian function to the PP map. We thus obtain values for the centre coordinates and standard deviations $\sigma_x$ and $\sigma_y$. We then calculate the EVF radius using the following formula (see Eq.~9 of \citealt{Rosolowsky_21}):
\begin{equation}
    R_{\rm EVF} = \eta \sqrt{\sigma_x \sigma_y} \,,
\end{equation}
where the factor $\eta=\sqrt{2\ln 2}=1.18$ comes by modelling the surface brightness as a two-dimensional Gaussian and using the half width at half maximum as the radius \citep{Rosolowsky_21}. We note that this procedure will overestimate the radius of unresolved EVFs, so the radii should be considered upper limits. 

Further, to obtain the galactocentric distance ($R_{\rm cen}$) of the EVFs, we use the coordinates of the EVF centre ($x$ and $y$), determined from the 2D Gaussian fit described above, in the formula
\begin{equation}
    R_{\rm cen}=\sqrt{(x-x_o)^2+(y-y_o)^2} \, ,
\label{eq:Rcen}
\end{equation}
where ($x_o$, $y_o$) are the coordinates of the galaxy centre. The measurement uncertainty of the EVF centre and the uncertainty in inclination is propagated to the uncertainty in $R_{\rm cen}$.

To estimate the velocity extent of the EVFs, we construct the total spectrum — intensity versus velocity — in the region enclosed by $R_{\rm EVF}$. We then fit a single Gaussian to the total spectrum. The standard deviation $\sigma_v$ of the fit is taken as a measure of the velocity dispersion, and the full width at half maximum (FWHM) of the fit as a measure of the velocity extent, $\Delta v = (2 \sqrt{2 \ln 2}) \sigma_v$.

Example spectra for the two EVFs identified in NGC\,1300 and Gaussian fits are shown in Fig.~\ref{fig:NGC1300_spectrum}. The uncertainties in all fitted parameters, including the centre coordinates, $\sigma_x$, $\sigma_y$ and $\Delta v$, were estimated from the square roots of the diagonal elements of the covariance matrix obtained during the respective fitting processes. These uncertainties reflect the confidence intervals in the fitting only and they do not take into account that the shape of the intensity distribution within EVF region might be non-Gaussian, that there might be contaminating emission from other features overlapping with the EVF in PP space, or that the radius can be overestimated when the EVF is unresolved.

To estimate the EVF masses, we calculate the total integrated CO intensity, $I_{\rm CO}$, within the determined $R_{\rm EVF}$. We then convert this total intensity into mass using the relation:
\begin{equation}
M_{\rm CO} = \alpha_{\rm CO} I_{\rm CO} \,,
\end{equation}
where $\alpha_{\rm CO} = 6.69 \ \mathrm{M_{\odot}\,pc^{-2}\,(K\,km\,s^{-1})^{-1}}$ is the CO(2-1)-to-H$_2$ conversion factor, and is derived from the Galactic value of 4.35 $\rm M_{\odot}pc^{-2}(K\,kms^{-1})^{-1}$ estimated by \citet{Bolatto_2013}, with the factor 0.70 accounting for the conversion between $^{12}$CO $J=2-1$ and $J=1-0$ transitions \citep{Leroy_22}. We adopt this procedure for simplicity, but note that it ignores galaxy-scale variations \citep[e.g.][]{denBrok2025} and might overestimate the mass for high-velocity dispersion features such as the EVF since high velocity dispersion often causes lower $\alpha_{\rm CO}$ values near galactic centres \citep{Teng2023}.

Following \citet{Bertoldi_McKee_1992} and \citet{Rosolowsky_21}, we calculate the virial masses and virial parameters as:
\begin{align}
M_{\rm vir} & = \frac{5 \sigma_v^2 R_{\rm EVF}}{G} \,, \\
\alpha_{\rm vir} & = \frac{2M_{\rm vir}}{M_{\rm CO}} \,.
\end{align}
The virial mass represents an estimate of the mass of the EVF under the assumption of dynamical equilibrium. The virial parameter $\alpha_{\rm vir}$ compares the virial mass to the CO-measured mass, and is a measure of the ratio between the kinetic and gravitational binding energy of the cloud. In dynamical equilibrium, the virial parameter is expected to be of order unity, while a large virial parameter indicates an out-of-equilibrium unbound situation (kinetic energy $\gg$ gravitational energy).

All estimated properties of the EVF regions are provided in Table \ref{tab:EVF_candidates}. 

\begin{table*}
\caption{Properties of the EVFs identified in this paper. Columns are: Galaxy ID, label of each EVF identified in each galaxy, galactrocentric distance from the EVF centre, radius of the EVF, velocity dispersion of the Gaussian fit to the EVF, velocity spread of the EVF (FWHM of the Gaussian fit), total CO(2--1) mass within each EVF region, the virial alpha parameter for each EVF and index for each feature which is used in figures to distinguish between EVFs.
}
\label{tab:EVF_candidates}
\begin{tabular}{@{}ccccccccc@{}}
\toprule
Galaxy ID & Feature  & $R_{\rm cen}$ & $R_{\rm EVF}$ & \textbf{$\sigma_v$} & $\Delta v$ & M$_{\rm CO(2-1)}$ & $\alpha_{\rm vir}$ & Index \\
&  & (kpc) & (kpc) & (\kms) & (\kms) & ($\rm \times10^5 M_{\odot}$) & \\ \midrule
NGC1097 & F1 & 3.53 $\pm$ 0.42 & 0.12 & 30.86 $\pm$ 0.39 & 72.68 $\pm$ 0.91 & 2.13 $\pm$ 0.61 & 1207 &1 \\
NGC1300 & F1 & 3.56 $\pm$ 0.23 & 0.10 & 13.33 $\pm$ 1.08 & 31.40 $\pm$ 2.55 & 0.39 $\pm$ 0.13 & 1043 &2\\
 & F2 & 4.77 $\pm$ 0.31 & 0.12 & 18.56 $\pm$ 0.29 & 43.70 $\pm$ 0.68 & 2.71 $\pm$ 0.64 & 368&3\\
NGC1512 & F1 & 3.00 $\pm$ 0.29 & 0.11 & 30.63 $\pm$ 1.05 & 72.13 $\pm$ 2.47 & 1.71 $\pm$ 0.44 & 1464 &4\\
NGC2566 & F1 & 3.09 $\pm$ 0.37 & 0.15 & 25.89 $\pm$ 0.60 & 60.97 $\pm$ 1.41 & 4.21 $\pm$ 1.27 & 550&5 \\
 & F2 & 3.56 $\pm$ 0.42 & 0.14 & 28.82 $\pm$ 0.70 & 67.87 $\pm$ 1.66 & 4.72 $\pm$ 1.33 & 565&6 \\
NGC2903 & F1 & 2.92 $\pm$ 0.37 & 0.09 & 30.24 $\pm$ 0.46 & 71.20 $\pm$ 1.08 & 1.88 $\pm$ 0.96 & 969&7 \\
 & F2 & 3.45 $\pm$ 0.44 & 0.07 & 18.81 $\pm$ 0.21 & 44.31 $\pm$ 0.50 & 1.21 $\pm$ 0.68 & 509&8 \\
NGC3627 & F1 & 1.08 $\pm$ 0.03 & 0.09 & 25.21 $\pm$ 0.27 & 59.37 $\pm$ 0.64 & 1.68 $\pm$ 0.51 & 778 &9\\
 & F2 & 2.79 $\pm$ 0.08 & 0.08 & 18.98 $\pm$ 0.19 & 44.71 $\pm$ 0.46 & 1.31 $\pm$ 0.45 & 524 &10\\
 & F3 & 3.60 $\pm$ 0.10 & 0.09 & 18.88 $\pm$ 0.36 & 44.47 $\pm$ 0.84 & 1.55 $\pm$ 0.55 & 492 &11\\
 & F4 & 1.37 $\pm$ 0.04 & 0.11 & 29.18 $\pm$ 0.42 & 68.72 $\pm$ 0.99 & 9.67 $\pm$ 2.61 & 223 &12\\
NGC4303 & F1 & 2.50 $\pm$ 0.17 & 0.14 & 11.50 $\pm$ 0.05 & 27.08 $\pm$ 0.13 & 5.55 $\pm$ 1.52 & 76 &13\\
 & F2 & 0.84 $\pm$ 0.06 & 0.13 & 17.91 $\pm$ 0.21 & 42.18 $\pm$ 0.50 & 7.11 $\pm$ 1.59 & 140 &14\\
NGC4535 & F1 & 1.17 $\pm$ 0.22 & 0.15 & 21.17 $\pm$ 0.63 & 49.85 $\pm$ 1.49 & 1.76 $\pm$ 0.68 & 903 &15\\
NGC4548 & F1 & 0.67 $\pm$ 0.06 & 0.14 & 21.05 $\pm$ 1.18 & 49.56 $\pm$ 2.78 & 0.50 $\pm$ 0.13 & 2925 &16 \\
 & F2 & 3.02 $\pm$ 0.25 & 0.14 & 15.73 $\pm$ 0.45 & 37.05 $\pm$ 1.05 & 0.86 $\pm$ 0.24 & 932 &17\\
NGC5643 & F1 & 0.73 $\pm$ 0.04 & 0.11 & 22.66 $\pm$ 0.25 & 53.36 $\pm$ 0.60 & 8.77 $\pm$ 1.79 & 146 & 18 \\
\bottomrule
\end{tabular}
\end{table*}

\subsection{Milky Way spectra}
\label{sec:MW_fitting}
As we view the Milky Way almost edge-on, we cannot separate the bar lanes in the PP maps as we did in Sect. \ref{sec:diagnostic} for 
external galaxies. Therefore in the MW, we identify both the EVF regions and the bar lanes in the PV diagram using the data from \citet{Bitran_1997}. As in Sect. \ref{sec:diagnostic}, for visualising the dust lanes and EVF regions we use the \texttt{glue} software. The dust lanes and EVF regions are then overplotted in different colours on top of the main emission in greyscale in the PP and PV maps in Fig.\,\ref{fig:MilkyWay}, where we highlight the EVF regions within magenta rectangular boxes. 
Figure~\ref{fig:MW_EVFspectra} presents the CO(1--0) spectra of the two EVF regions. The left panel presents the total spectrum within the extracted EVF region at $l=3.2^\circ$, corresponding to emission at a longitude range 2.7--3.4\degr, latitude not restricted. The right panel presents the total spectrum within the extracted EVF region at $l=5.4^\circ$, corresponding to the emission at a longitude range of 5.0--5.6\degr, latitude unresricted. Due to our embedded perspective
within the MW disk, the EVFs are “contaminated” by over-
lapping features along the line-of-sight, resulting in spectra
that appear more complex than those of external galaxies.

The spectrum on the left panel (EVF at $l=3.2^\circ$) shows contamination from a feature that is known to be a secondary dust lane \citep{Liszt_2008}, which appears as a narrow peak superimposed on the broader EVF peak. Since the single Gaussian fitting technique used for external galaxies is inadequate for cases with overlapping features, we simply adopt a double Gaussian fitting approach to accurately model multiple components and account for foreground contamination. This yields a FWHM $70.64 \pm 2.07$\,\kms~for the EVF component and a $13.92 \pm 2.14$\,\kms~for the component of the overlapping dust lane. We take the first as our estimate of the EVF velocity extension $\Delta v$.

The spectrum on the right panel (EVF at $l=5.4^\circ$) shows two broad peaks, and it is not obvious what constitutes foreground contamination and what is genuine EVF emission. To produce nevertheless an estimate of $\Delta v$, we again fit a double Gaussian and consider the broadest peak to be the EVF, resulting in $\Delta v = 68.40 \pm 4.19$\,\kms~.

We note that the method for estimating $\Delta v$ for the MW and external galaxies is slightly different as we used a double Gaussian to get rid of contaminating features in the first case. This difference in methodology should be borne in mind when comparing $\Delta v$, but it does not affect any of our conclusions below as they are not very sensitive to the exact values of $\Delta v$.

\begin{figure*}
    \hspace{-0.5cm}
    \includegraphics[width=0.517\textwidth]{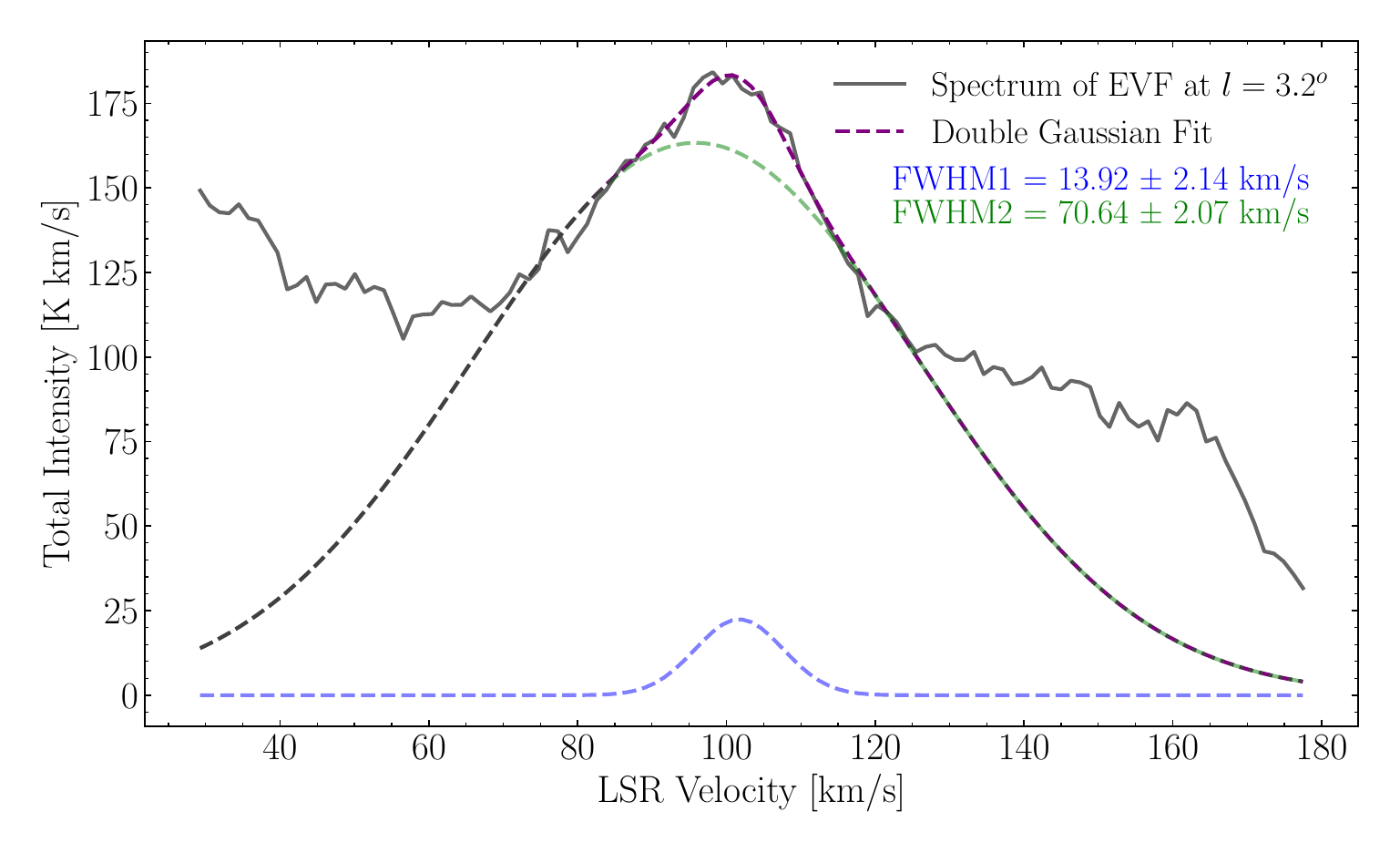}\hspace{-0.3cm}
    \includegraphics[width=0.517\textwidth]{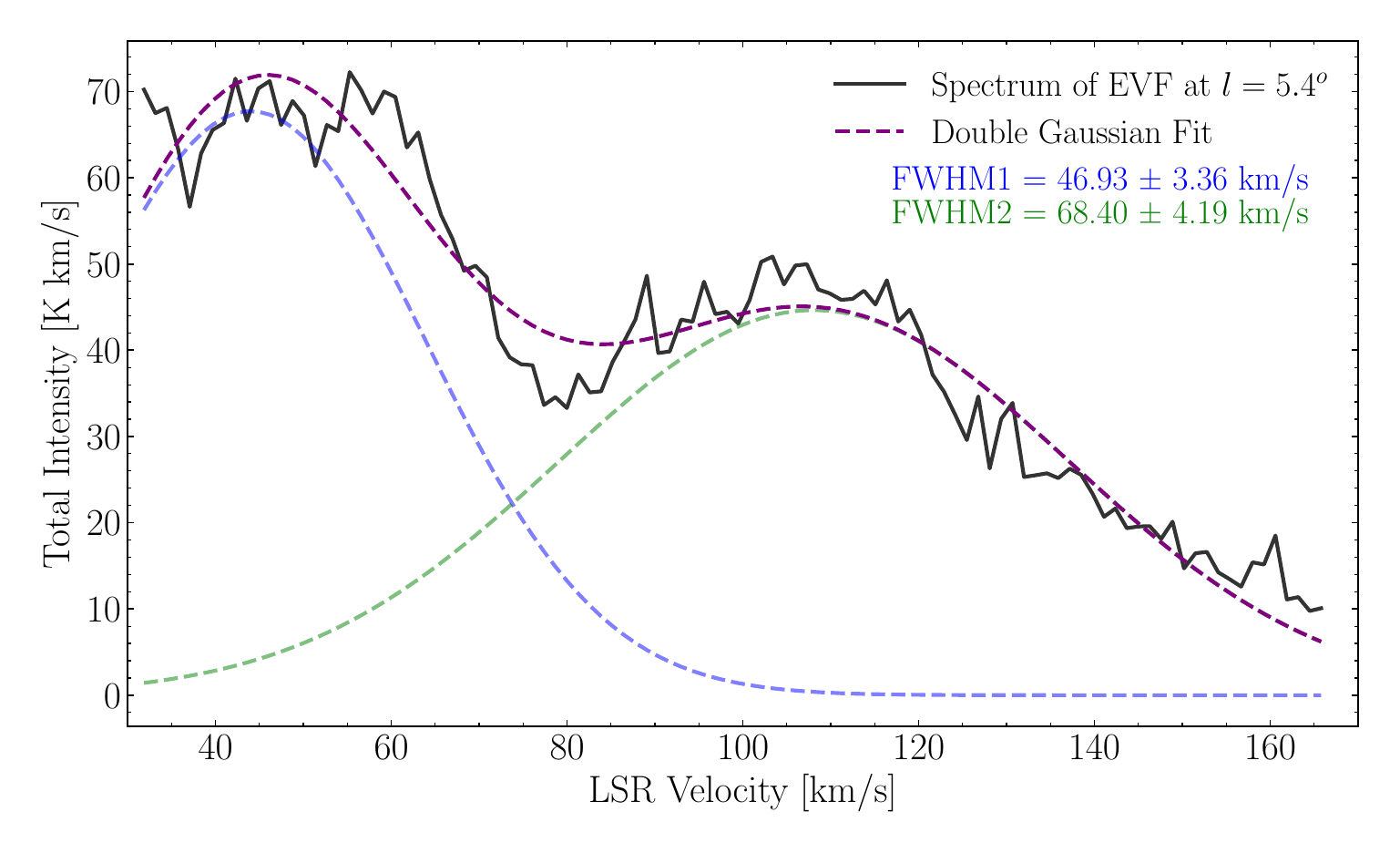}
    \caption{The spectra (black full curves) and double Gaussian fits (dashed curves) for two MW EVFs highlighted in Fig.~\ref{fig:MilkyWay} and located at $l=3.2^\circ$ (left) and $l=5.4^\circ$ (right). The two components of the double Gaussian fits are in blue and green, and their sum is in purple.}
    \label{fig:MW_EVFspectra}
\end{figure*}

\section{Results}
\label{sec:results}

After examining the PP and PV diagrams of the 29 galaxies in our sample, we identified EVFs in 10 out of 29 galaxies, which represents approximately 34\% of the total. The remaining 19 galaxies (about 66\%) did not show obvious features that we could identify as EVFs, although it is not excluded (and is actually likely) that higher resolution/sensitivity surveys might reveal more features. The 10 galaxies exhibiting signatures of EVFs are: NGC\,1097, NGC\,1300, NGC\,1512, NGC\,2566, NGC\,2903, NGC\,3627, NGC\,4303, and NGC\,4535,  NGC\,4548 and NGC\,5643. Amongst these 10, 4 galaxy exhibited one EVF each, while the remaining 6 galaxies exhibited more than one EVF. In total, we identified 18 EVFs within 10 galaxies. To interpret the nature of these features, we study their spectra, locations and physical properties as discussed below.

Let us use NGC\,1300, shown in Figs.~\ref{fig:NGC1300}, as an illustrative example. In this galaxy, we identified two EVFs: one on the dust lane coloured in red at a radius of approximately 3.6\,kpc from the galactic centre, and one on the dust lane coloured in blue at a radius of approximately 4.8\,kpc. These are marked with rectangular boxes in Fig.~\ref{fig:NGC1300}. The EVFs show enhanced CO(2--1) intensity in the PP diagrams, confined to rather round, blob-like regions. The most prominent of the two EVFs is the feature on the dust lane coloured in blue. Similar to EVFs observed in the MW, this feature has a large vertical and narrow horizontal extent in the PV diagram. 

In the vicinity of approximately 12 of the EVF regions, while CO(2--1) emission appears as round blob-like features, the emission in 7.7\,$\rm \mu m$ is resolved into smaller clumps. An example of this can be seen in Fig.~\ref{fig:JWST-Zoom-NGC1300} for NGC\,1300, see also Figs.~\ref{fig:NGC1097_b}, \ref{fig:NGC2566_b}(F2), \ref{fig:NGC2903_b}, \ref{fig:NGC3627_b}, \ref{fig:NGC4303_b}(F1) for several other galaxies. These clumps might contain enhanced star formation activity in smaller collapsed regions, possibly triggered by molecular cloud collisions near the EVFs. In fact, a recent study of NGC\,3627 by \citep{Maeda_2025} shows that, despite being less efficient at high collisional velocities, cloud collisions can still enhance the star formation in bar lanes. For the remaining 6 EVF regions, we do not see strong intensity enhancement or clumpiness in 7.7\,$\rm \mu m$ emission; see Figs.~\ref{fig:NGC1512_b} and \ref{fig:NGC4548_b} as examples. 

To check whether unbarred galaxies can also contain EVFs, we examined all galaxies that are classified as lacking bar-like features in the molecular gas morphology (Class A) in \citet{Stuber_2023} which are also classified as unbarred (`SA') in photometric studies \citep{deVaucouleurs_1991,Buta_2015}. These 13 galaxies are presented in Table \ref{tab:PHANGS_galaxysample_unbarred}. Of these, only one galaxy --NGC\,628-- exhibited a tentative EVF-like signature in the PV diagram, along one of its spiral arms. The CO velocity and velocity dispersion maps of NGC\,628 appear relatively undisturbed, but several faint gas streams are visible, seemingly connecting to the region where the EVF occurs. These findings suggest that a cloud collision might be happening in the spiral arm of this galaxy, leading to the emergence of EVF signatures in their PV diagrams. However, the general absence of EVF signatures in unbarred galaxies supports the argument that EVFs arise primarily due to cloud-cloud or stream-stream collisions in bar lanes.

\begin{table}
\centering
\caption{Sample of galaxies classified consistently as `unbarred' from both CO morphology \citep{Stuber_2023} and photometry \citep{deVaucouleurs_1991,Buta_2015} used in this work to search for EVFs as test cases. Columns are: Galaxy ID; Distances obtained from \citet{Anand_2021}; Physical resolution in native resolution cube \citet{Leroy_2021}; kinematic position angle of LON and inclination inferred from \citet{Lang_2020}; Presence of EVFs identified in this work, based on PP and PV diagrams, as discussed in Sect.\ref{sec:results}.`Y' indicates that EVF signatures are found in the galaxy, while `N' indicates no clear signatures of EVFs are detected.}
\hspace{-0.5cm}
\resizebox{1.04\columnwidth}{!}{%
\begin{tabular}{lccccc}
\hline
{Galaxy ID} & Distance &  Resolution &  PA of LON & $i$ &  EVF \\ & (Mpc) & (pc) & (deg) & (deg) & (Y,N) \\
\hline
NGC\,0628 & 9.84 & 53.5  & 21   & 9  & Y \\
NGC\,1546 & 17.69 & 109.6 & 148  & 70 & N \\
NGC\,1792 & 16.20 & 150.9 & 134  & 65 & N \\
NGC\,1809 & 19.95 & 136.0 & 139  & 58 & N \\
NGC\,2090 & 11.75 & 73.8  & 13   & 65 & N \\
NGC\,3137 & 16.37 & 120.0 & 180  & 70 & N \\
NGC\,3521 & 13.24 & 85.5  & 163  & 69 & N \\
NGC\,3596 & 11.30 & 66.9  & 78   & 25 & N \\
NGC\,3621 & 7.06  & 62.4  & 164  & 66 & N \\
NGC\,4254 & 13.10 & 113.1 & 68   & 34 & N \\
NGC\,4298 & 14.92 & 114.7 & 134  & 59 & N \\
NGC\,4689 & 15.00 & 86.0  & 164  & 39 & N \\
NGC\,4826 & 4.41  & 26.9  & 114  & 59 & N \\
\hline
\end{tabular}}

\label{tab:PHANGS_galaxysample_unbarred}
\end{table}

\subsection{Physical properties of EVFs} \label{sec:EVFproperties}

\begin{figure*}
    \hspace{-0.45cm}
 \includegraphics[width=0.9\textwidth]{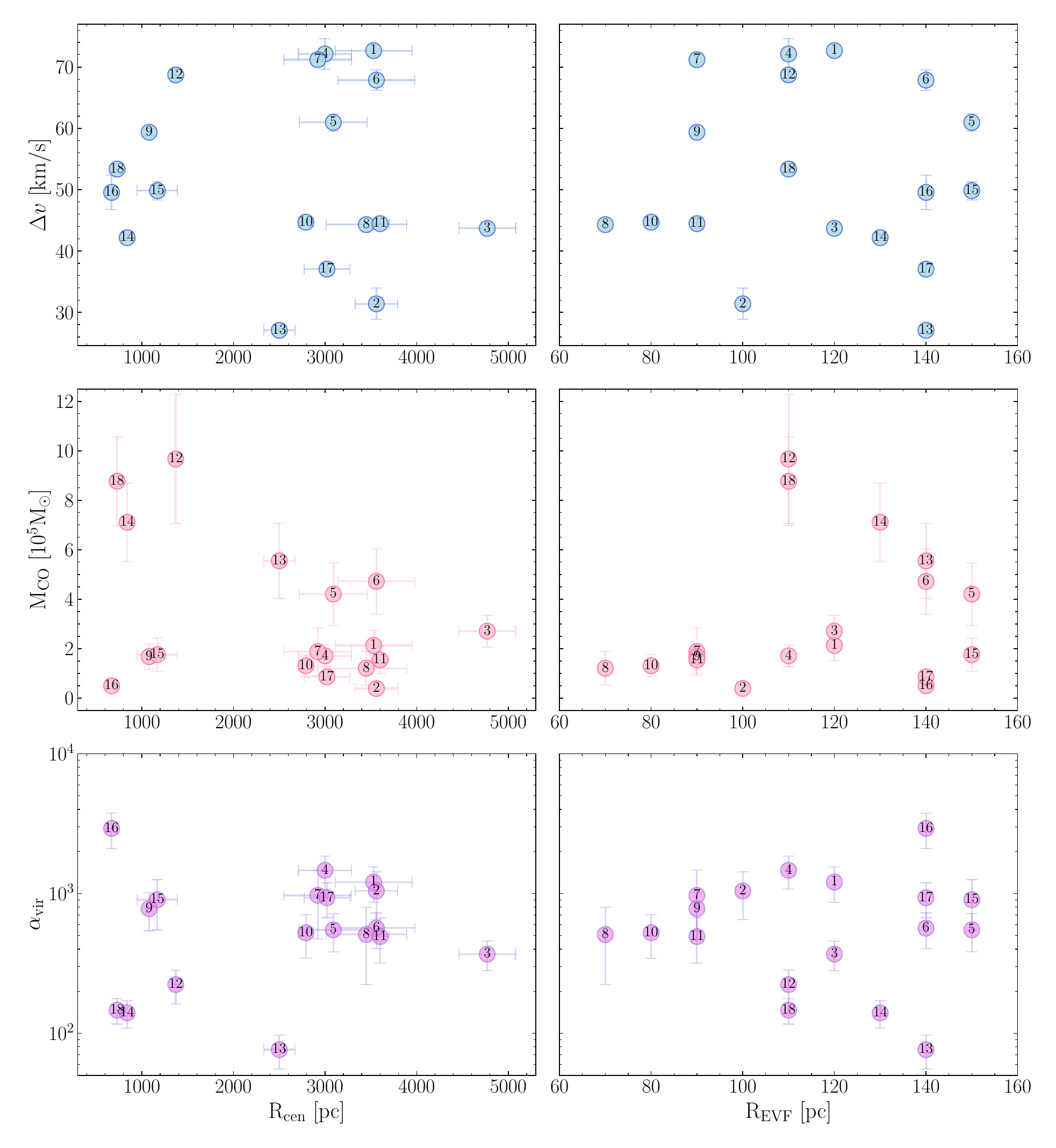}
    \caption{Physical properties of EVFs plotted against the radius from the galaxy center (left) and the physical radius of the EVF (right). From top to bottom: velocity extent $\Delta v$, mass $M_{\rm CO}$, virial parameter $\alpha_{\rm vir}$. Each point is numbered according to the assigned EVF index listed in Table \ref{tab:EVF_candidates}.}
    \label{fig:EVF_subplots}
\end{figure*}

Table~\ref{tab:EVF_candidates} lists the properties of all of the EVFs identified in our sample. Typical masses are in the range $10^5 \mhyphen 10^6\, M_\odot$, while typical sizes are of the order of 100 pc. These values are similar to the MW EVFs shown in Fig.~\ref{fig:MilkyWay}, which have radii of $\sim 100$~pc and estimated masses ranging from a few $10^4\, M_\odot$ to a few $10^6\, M_\odot$ depending on the method used \citep{Liszt_2006,Gramze_2023}. Figure~\ref{fig:EVF_subplots} shows the EVFs' properties against their galactocentric radius (left panels) and their physical radius (right panels), illustrating that they are found at different distances from the galactic centre, spanning the full extent of the bar lanes. All these characteristics are consistent with the EVFs identified in our sample being analogues of the MW features. 

For all of the EVFs we identified, the virial parameter, $\alpha_{\rm vir}$, is exceptionally large, ranging from a few hundred to several thousand (see Table~\ref{tab:EVF_candidates}). Bound molecular clouds should have $\alpha_{\rm vir}\leq2$, while unbound systems have $\alpha_{\rm vir}>2$ \citep[e.g.][]{Bertoldi_McKee_1992}. This indicates that EVFs are all unbound systems that are very far from virial equilibrium, consistent with the idea that they are the sites of extreme cloud-cloud (or stream-stream) collisions. We caveat that if the EVFs are unresolved in the PHANGS-ALMA CO data (which is possible given their `blobby' appearance), then we will overestimate their virial parameter since $R_{\rm EVF}$ will be larger than the true cloud size. If instead, we overestimate $M_{\rm CO}$ due to the simple $\alpha_{\rm CO}$ prescription adopted (Sect.~\ref{sec:properties_estimation}), then we will underestimate the virial parameter. However, these effects are unlikely to affect our conclusions, because even in the worst case we would not expect the virial parameters to change by more than a factor of a few, so the EVFs would still be strongly super-viral.

The formulae for both the virial mass and the virial parameter, given in Sect.~\ref{sec:properties_estimation}, which result in the values given above, assume an isotropic velocity distribution. On the other hand, if the EVFs are sites of collisions of clouds moving with markedly different velocities, their velocity extent may be dictated by the velocity difference between the colliding clouds. In particular, it may not be isotropic, as the colliding clouds move in the galaxy plane. If the velocity dispersion is not isotropic, but mainly in the disc plane, one should observe a trend of the EVF velocity difference being larger in galaxies that are more highly inclined. 

In the left panel of Fig.~\ref{fig:sinalpha_dv}, we show the velocity extent, $\Delta v$ against $\sin(i)$ for our EVFs. We include the MW results for comparison, although we remind the reader that the method for obtaining the $\Delta v$ is slightly different (see Sect.~\ref{sec:MW_fitting}) than how we inferred the parameter in external galaxies. For external galaxies, $\Delta v$, ranges from approximately 27\,\kms~to 72\,\kms. Both of the MW EVFs with $\sim$70\kms\ are on the higher end of this range. The figure shows a rough correlation between $\Delta v$ and $\sin(i)$, consistent with expectations for the EVF velocity extent resulting from velocity difference of the colliding clouds. Particularly, it suggests that the $\Delta v$ is likely not driven by isotropic velocity dispersion. Although the scatter in this figure is large, this is not surprising since in different galaxies, and in different EVFs clouds can collide with different relative velocities.

If the velocity extent of EVFs is dictated entirely by the velocity difference of colliding clouds, then one should expect the largest velocity extent for bars roughly perpendicular to the line of nodes. This is because, to the first approximation, the radial velocity in the bar lane gets reverted (outflow upstream from the bar lane turns into inflow downstream from it), while the azimuthal velocity continues as prograde rotation, and the radial velocity dominates the observed line-of-sight velocity in the direction perpendicular to the line of nodes.
In the right panel of Fig.~\ref{fig:sinalpha_dv}, we show the $\Delta v/\sin(i)$ as a function of $\rm PA_{LON} - PA_{bar}$. There is no clear minimum at $\rm PA_{LON} = PA_{bar}$. This may indicate that turbulent random motions resulting from the collision significantly contribute to the velocity extent of EVFs, which therefore is not solely caused by the velocity difference of the colliding clouds.

\begin{figure*}
        \centering
\includegraphics[width=1\textwidth]{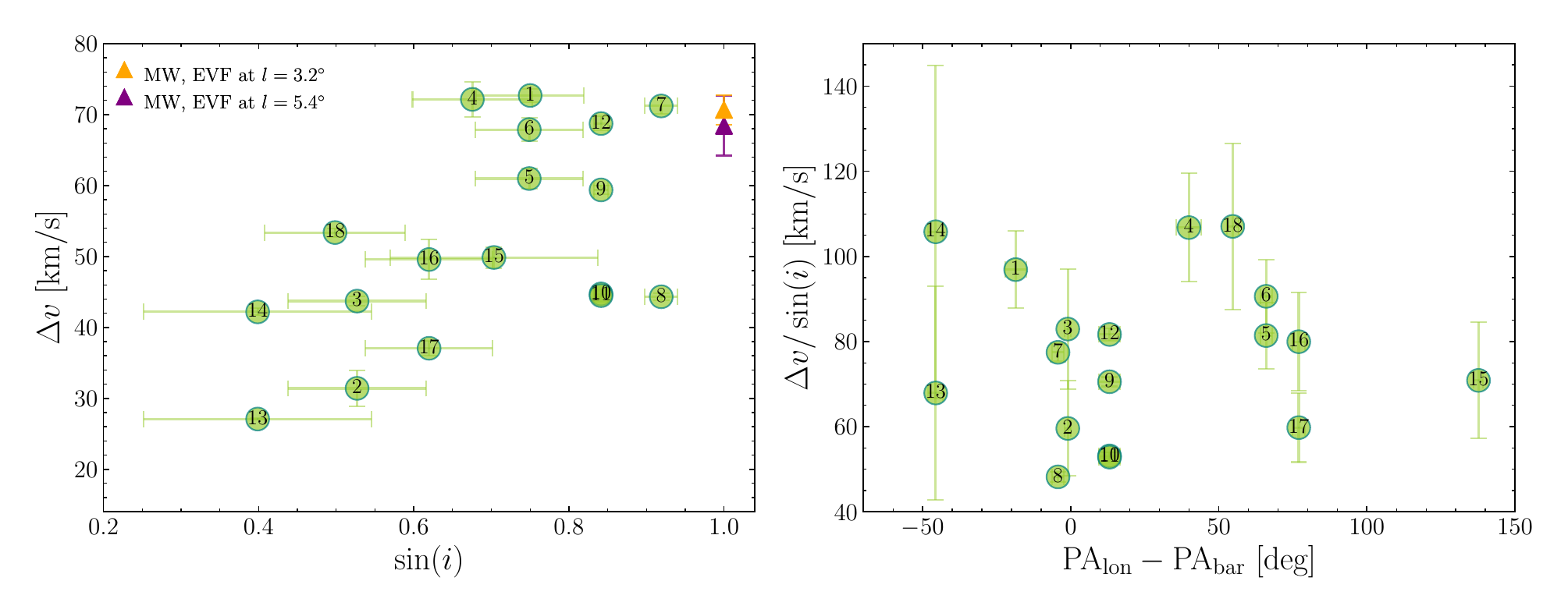}
    \caption{The distribution of $\Delta v$ against $\sin(i)$ (left) and the distribution of $\Delta v$/$\sin(i)$ against the difference of PA of the LON and PA of the bar (right). The points are numbered according to galaxy and EVF index, as listed in Table \ref{tab:EVF_candidates}. Triangle markers in left panel represent the measurements from the EVF of MW. 
    }
    \label{fig:sinalpha_dv}
\end{figure*}

\subsection{JWST gas streams}
\label{results:EVFlocaitons} 

PHANGS-JWST observations have no velocity information, but are much more sensitive to low surface density gas and have higher spatial resolution than the PHANGS-ALMA CO data. We use them to trace the gas that falls below the sensitivity threshold of the CO data. The 7.7\,$\rm \mu m$ band observations of NGC\,1300, presented in the bottom panel of Fig~\ref{fig:NGC1300}, reveals many streams of gas in the bar region surrounding the dust lanes that indeed are invisible in the CO data. Several streams connecting to the EVFs are marked by faint magenta arrows. These streams appear to extend continuously from the opposite side of the galaxy for several kpc, either in straight or slightly curving forms, and reach the location of the EVF, where enhanced CO(2--1) intensity is seen. In the cloud-cloud collision interpretation (see Sect.~\ref{sec:intro}), EVFs originate from collisions of gas streams along the bar lanes with gas streams coming from the opposite side of the galaxy. We hypothesise that these are the streams that collide with the dust lane streams, with almost opposite velocity vectors, giving rise to the EVFs. The direction of the streams shown by the magenta arrows is an educated guess based on the observed morphology and theoretical knowledge of gas flow in barred potentials.

Upon examining the 7.7\,$\rm \mu m$ band observations of all galaxies with EVFs, we find that out of 18 EVF regions, 9 are associated with faint streams extending from the opposite end of the bar toward the EVF location. In some galaxies, such as NGC\,1300, multiple faint streams reach the EVF regions. The galaxies displaying these streams of overshot gas are NGC\,1300, NGC\,1097, NGC\,2566 (only F1), NGC\,3627 (except F1), NGC\,4548 (F2), and NGC\,5643. For 5 EVFs ($\sim$28\%) we cannot determine whether streams are associated due to star-forming activity within the bar regions. These cases are found in NGC\,2566 (F2), NGC\,2903, NGC\,3627 (F1), and NGC\,4303 (F1). For the remaining 4 ($\sim$22\%) EVFs— in NGC\,1512 (F1), NGC\,4303 (F2),  NGC\,4535 (F1), and in NGC\,4548 (F1)—we did not identify any associated streams of overshot gas. 

Several factors may contribute to the absence of observable streams in other galaxies exhibiting EVFs. One possibility is an intrinsically low gas density, where the emission is too weak to be detected in 7.7 $\rm \mu m$  observations. Additionally, environmental effects such as global rotational flow and turbulence within the intergalactic medium could introduce kinematic disturbances, leading to gas dissipation through mixing, making the streams too faint to be traced. Stacking observations across different MIRI filters could help produce deeper images and reveal fainter structures. However, since such data were not consistently available for all galaxies, we did not pursue this approach in our analysis. It is also possible that that the stream was a short segment which is lost to the collision.

In the collision interpretation, the geometry of the collision might lead us to expect that the EVFs should occur on the trailing side of the bar lanes. Although it is difficult to assess this conclusively as dust lanes are thin and the spatial resolution is not sufficient to examine this in detail, some galaxies appear consistent with this expectation (e.g.\  NGC\,1512, NGC\,2903, NGC\,3627), but for some other galaxies the EVFs seem to occur on the leading side of the bar lanes (e.g.\ NGC\,1097 and NGC\,4535). There might be various explanations for this phenomenon within our cloud-cloud collision hypotheses. For example, the gas could traverse the dust lane while the collision is still taking place, as the simulations show that it is a collision between streams, which happens over an extended period in time \citep{Sormani_2019}. Another reason might be the gas clumpiness, so the incoming streams might find holes in the gas distribution and only collide with other gas parcels when it has traversed the relatively short thickness of the dust lane (in other words, the mean free path might be comparable to the lane thickness).

\section{Summary \& Conclusions} 
\label{sec:conclusion}

The inner regions of the MW are known to harbour an enigmatic population of prominent molecular clouds characterised by extremely broad lines. The physical origin of these ``extended velocity features'' (EVF) has remained controversial, but a connection with the dust lanes of the Galactic bar has been suggested by several authors.

In this study, we searched for analogous features in the bar dust lanes of 29 nearby barred galaxies using the PHANGS-ALMA CO(2--1) survey, with the aim of confirming the existence of EVF-like features in other galaxies and of using the external perspective to gain insight into their origin. We found EVFs in 10 out of the 29 galaxies, corresponding to 34\% of our sample. The EVF features stand out as relatively compact structures that have an unusually large velocity spread in the PPV diagrams compared to the rest of the material on the dust lane. Upon examining a test sample of 13 unbarred galaxies, we find a general absence of EVFs in their PPV diagrams.

We analysed the physical properties of the EVFs and found that they all have virial parameters ranging from few 100s to several 1000s, indicating highly unbound structures. The most likely explanation for their origin is extreme cloud-cloud, or more appropriately stream-stream, collisions with relative velocities in excess of 100\kms. Further support to this interpretation is provided  by PHANGS-JWST  MIRI 7.7$\mu m$ observations, revealing low-intensity, CO-invisible streams of gas in 50\% of the EVF cases, that appear, based on their morphology, to be hitting the dust lanes at the location where EVFs are found.

Our findings support the idea that EVFs are the result of extreme collisions between gas streams occurring in the dust lanes of barred galaxies. They are probably the clearest examples of cloud-cloud collisions available in the literature. Thus, they are unique astrophysical laboratories to study the physics and chemistry of complex cloud collisions and their impact on star formation in galaxies.

\section*{Acknowledgements}
TK acknowledges the joint studentship support from Liverpool John Moores University the Faculty of Engineering and Technology, and the Science Technology Facilities Council. TK also acknowledges the financial support from The Leverhulme Trust. MCS acknowledges financial support from the European Research Council under the ERC Starting Grant ``GalFlow'' (grant 101116226) and from Fondazione Cariplo under the grant ERC attrattivit\`{a} n. 2023-3014. FF is supported by a UKRI Future Leaders Fellowship (grant no. MR/X033740/1). MC gratefully acknowledges funding from the DFG through an Emmy Noether Research Group (grant number CH2137/1-1). COOL Research DAO is a Decentralized Autonomous Organization supporting research in astrophysics aimed at uncovering our cosmic origins. SCOG and RSK acknowledge financial support from the European Research Council via the ERC Synergy Grant ``ECOGAL'' (project ID 855130) and from the Heidelberg Cluster of Excellence (EXC 2181 - 390900948) ``STRUCTURES'', funded by the German Excellence Strategy. RSK also acknowledges support from the German Ministry for Economic Affairs and Climate Action in project ``MAINN'' (funding ID 50OO2206). RSK is grateful for computing resources provided by the Ministry of Science, Research and the Arts (MWK) of the State of Baden-W\"{u}rttemberg through bwHPC and the German Science Foundation (DFG) through grants INST 35/1134-1 FUGG and 35/1597-1 FUGG, and also for data storage at SDS@hd funded through grants INST 35/1314-1 FUGG and INST 35/1503-1 FUGG. RSK also thanks the 2024/25 Class of Radcliffe Fellows for their company and for highly stimulating discussions. FP acknowledges support from the Horizon Europe research and innovation programme under the Marie Skłodowska-Curie grant “TraNSLate” No 101108180, and from the Agencia Estatal de Investigación del Ministerio de Ciencia e Innovación (MCIN/AEI/10.13039/501100011033) under grant (PID2021-128131NB-I00) and the European Regional Development Fund (ERDF) ``A way of making Europe''. 

MQ acknowledges support from the Spanish grant PID2022-138560NB-I00, funded by MCIN/AEI/10.13039/501100011033/FEDER, EU.
\\

\noindent This paper makes use of the following ALMA data: \\
ADS/JAO.ALMA\#2012.1.00650.S, \\
ADS/JAO.ALMA\#2013.1.00803.S, \\ 
ADS/JAO.ALMA\#2013.1.01161.S, \\
ADS/JAO.ALMA\#2015.1.00121.S, \\ 
ADS/JAO.ALMA\#2015.1.00782.S, \\
ADS/JAO.ALMA\#2015.1.00925.S, \\
ADS/JAO.ALMA\#2015.1.00956.S, \\  
ADS/JAO.ALMA\#2016.1.00386.S, \\ 
ADS/JAO.ALMA\#2017.1.00392.S, \\
ADS/JAO.ALMA\#2017.1.00766.S, \\ 
ADS/JAO.ALMA\#2017.1.00886.L, \\ 
ADS/JAO.ALMA\#2018.1.00484.S, \\
ADS/JAO.ALMA\#2018.1.01321.S, \\  
ADS/JAO.ALMA\#2018.1.01651.S, \\ 
ADS/JAO.ALMA\#2018.A.00062.S, \\
ADS/JAO.ALMA\#2019.1.01235.S, \\  
ADS/JAO.ALMA\#2019.2.00129.S, \\

ALMA is a partnership of ESO (representing its member states), NSF (USA) and NINS (Japan), together with NRC (Canada), MOST and ASIAA (Taiwan), and KASI (Republic of Korea), in cooperation with the Republic of Chile. The Joint ALMA Observatory is operated by ESO, AUI/NRAO and NAOJ. The National Radio Astronomy Observatory is a facility of the National Science Foundation operated under a cooperative agreement by Associated Universities, Inc.

Part of this work is based on observations made with the NASA/ESA/CSA JWST. The data were obtained from the
Mikulski Archive for Space Telescopes at the Space Telescope Science Institute, which is operated by the Association of Universities for Research in Astronomy, Inc., under NASA contract NAS 5-03127 for JWST. These observations are associated with program 2107. 

\textit{Software:} This research made use of \texttt{ASTROPY} and its affiliated packages \citep{Astropy_2018, Astropy_2022}; and \texttt{glue} software of multidimensional data visualisation  \citep{Beaumont_2015, Robitaille_2017}.

\section*{Data Availability}

The specific observations used in this work are obtained from PHANGS-ALMA, publicly accessible at \url{https://www.canfar.net/storage/list/phangs/RELEASES/PHANGS-ALMA/}, and the  PHANGS-JWST Treasury Survey, available at DOI: \url{http://dx.doi.org/10.17909/ew88-jt15}. As a continuation of Cycle 1, the PHANGS-JWST Treasury Survey Cycle 2 with program ID 3707 (PI: Leroy) is ongoing, and a publication detailing the data is currently in preparation.

\label{lastpage}

\bibliographystyle{mnras}
\bibliography{main} 
\appendix
\counterwithin{figure}{section}

\section{Diagnostic maps, images and plots of sample galaxies which are exhibiting EVFs}
\label{sec:app_diagnostics_sample}

The figures in this section are configured as follows.

For galaxies where EVFs are identified, figures with caption numbering ending with an ``a"  (e.g., Figs.~\ref{fig:NGC1097}-- \ref{fig:NGC5643}) present the CO(2--1) PP and PV, JWST/MIRI 7.7\,$\rm \mu m$ band image, and spectra with Gaussian fits within the identified EVF regions. The allocation of the panels are configured similarly to Figs.~\ref{fig:NGC1300} and \ref{fig:NGC1300_spectrum}. Following these, figures with caption numbering ending with a ``b" (e.g., Figs.~\ref{fig:NGC1097_b} -- \ref{fig:NGC5643_b}) provide the zoom-in view of the regions marked with rectangular boxes in the 7.7\,$\rm \mu m$ band observations of the corresponding galaxy similar to what is shown in Fig.~\ref{fig:JWST-Zoom-NGC1300}.

\renewcommand{\thefigure}{A1\alph{figure}}

\setcounter{figure}{0} 

\begin{figure*}
    \centering
    \includegraphics[width=1\textwidth]{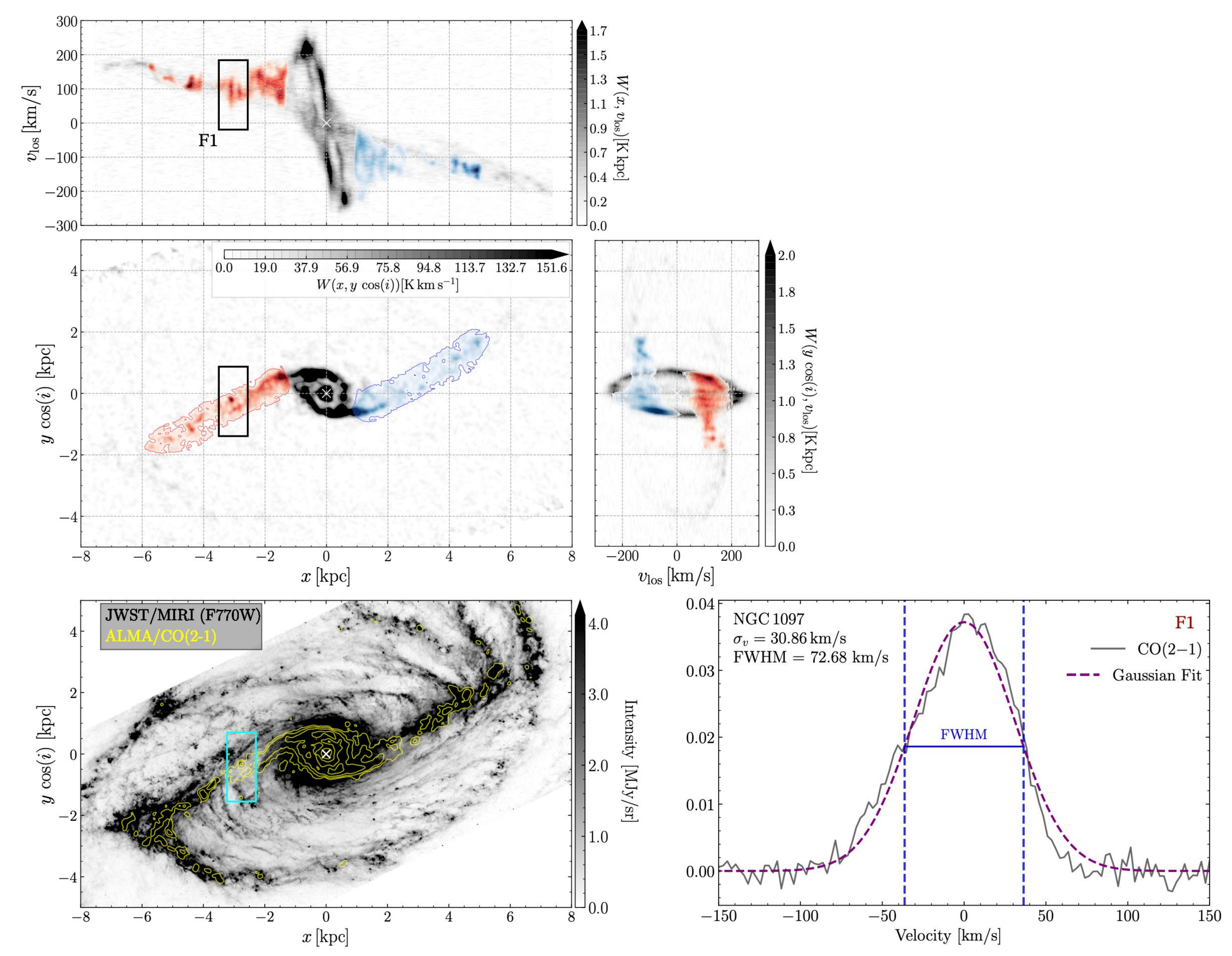}
    \caption{Diagnostic maps, 7.7\,$\rm \mu m$ band observations and spectra within EVF regions for NGC\,1097.}
    \label{fig:NGC1097}
\end{figure*}

\begin{figure*}
    \centering
    \includegraphics[width=0.35\textwidth]{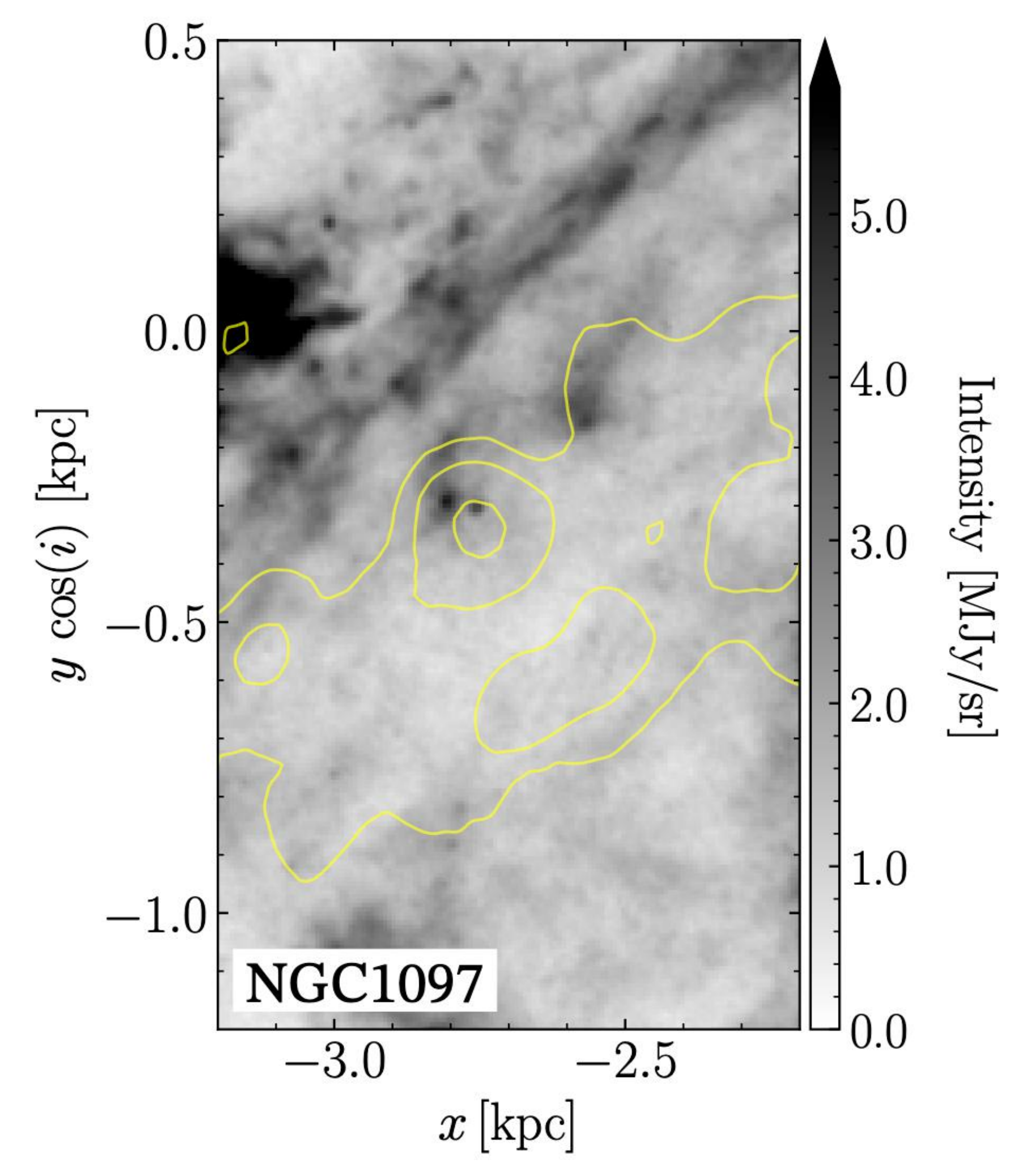}
    \caption{Zoomed-in 7.7\,$\rm \mu m$ band observations of NGC\,1097.}
    \label{fig:NGC1097_b}
\end{figure*}

\renewcommand{\thefigure}{A2\alph{figure}}

\setcounter{figure}{0} 

\begin{figure*}
    \centering
    \includegraphics[width=1\textwidth]{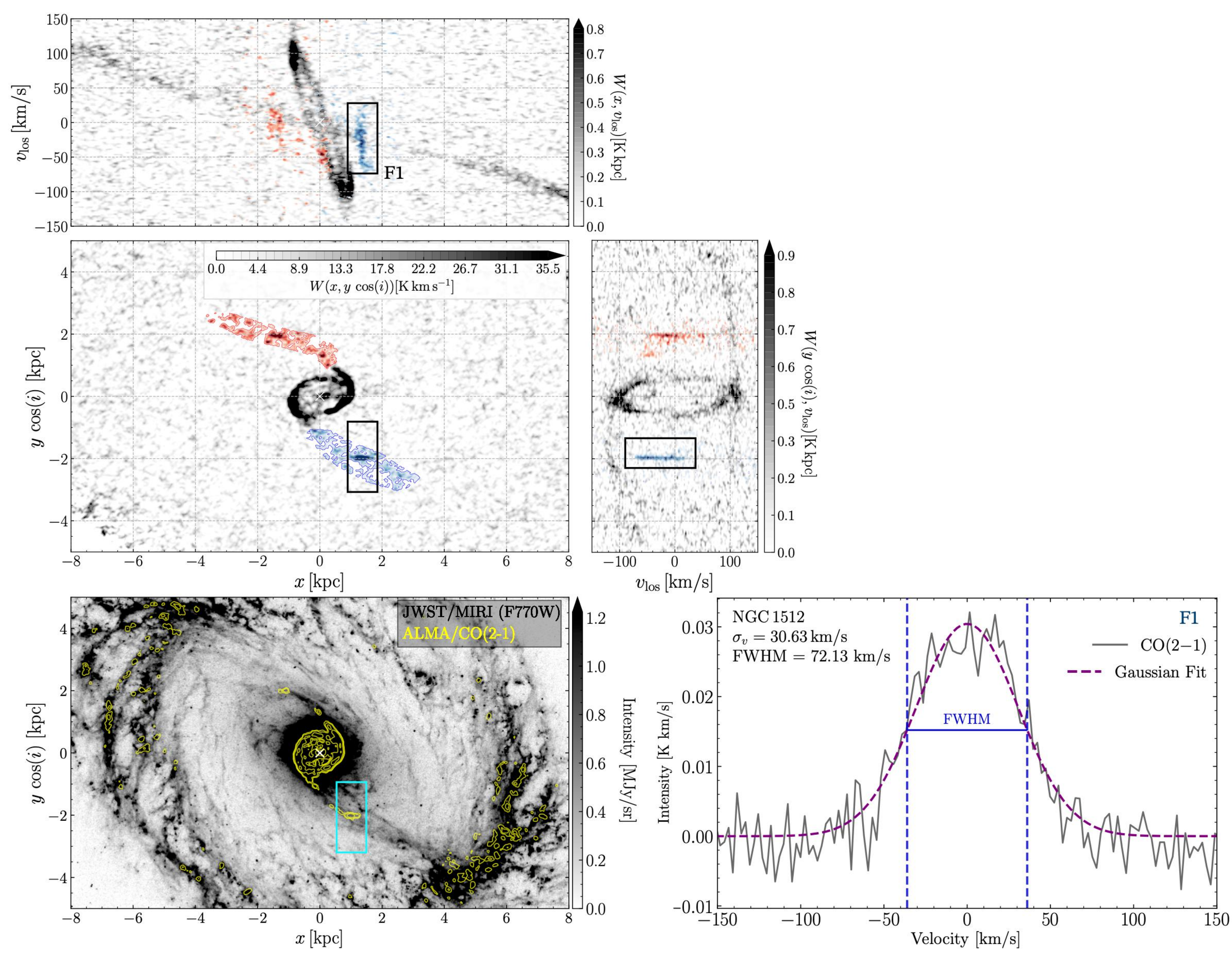}
    \caption{Diagnostic maps, 7.7\,$\rm \mu m$ band observations and spectra within EVF regions for NGC\,1512.}
    \label{fig:NGC1512}
\end{figure*}

\begin{figure*}
    \centering
    \includegraphics[width=0.35\textwidth]{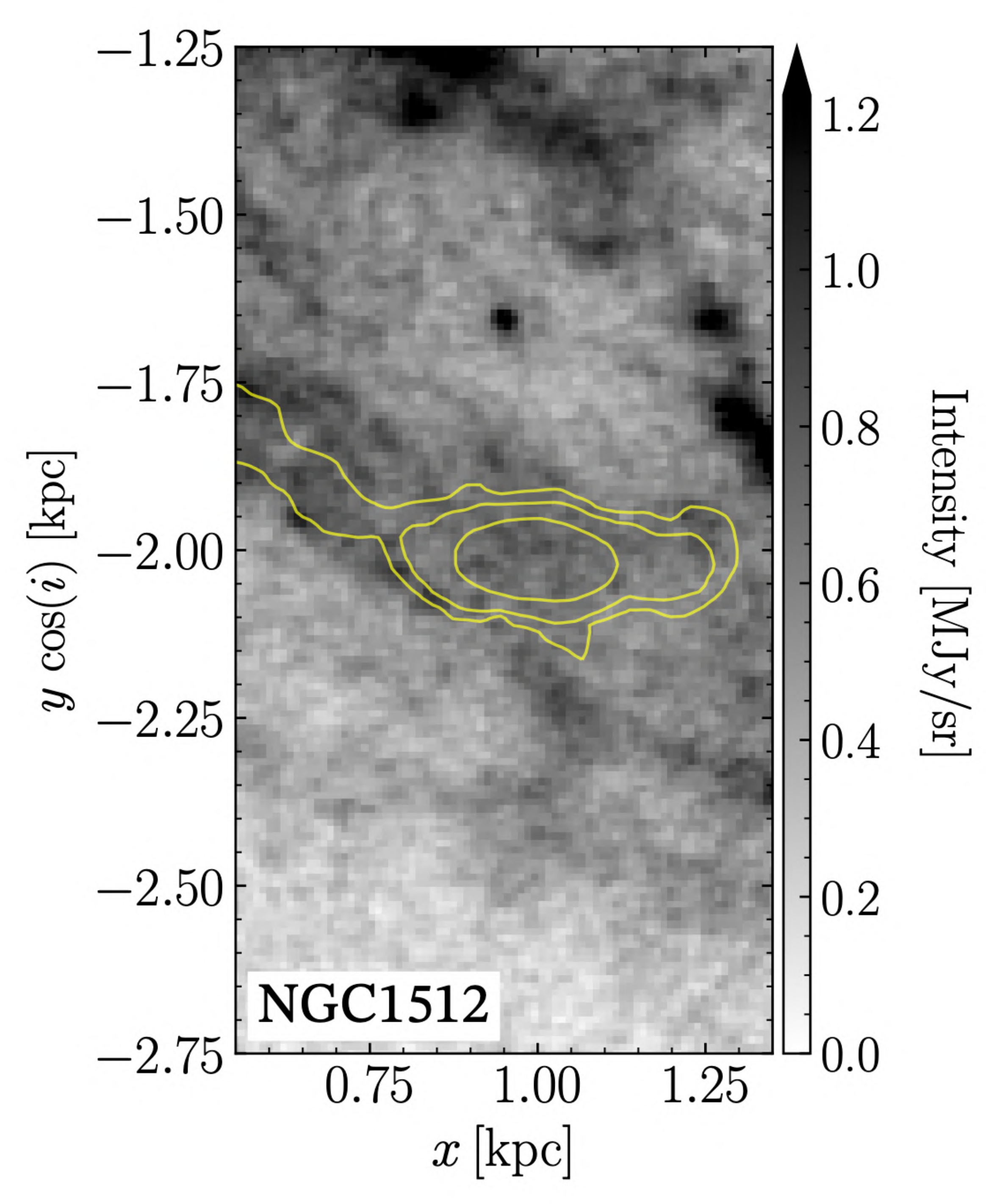}
    \caption{Zoomed-in 7.7\,$\rm \mu m$ band observations of NGC\,1512.}
    \label{fig:NGC1512_b}
\end{figure*}

\renewcommand{\thefigure}{A3\alph{figure}}
\setcounter{figure}{0} 

\begin{figure*}
    \centering    \includegraphics[width=1\textwidth]{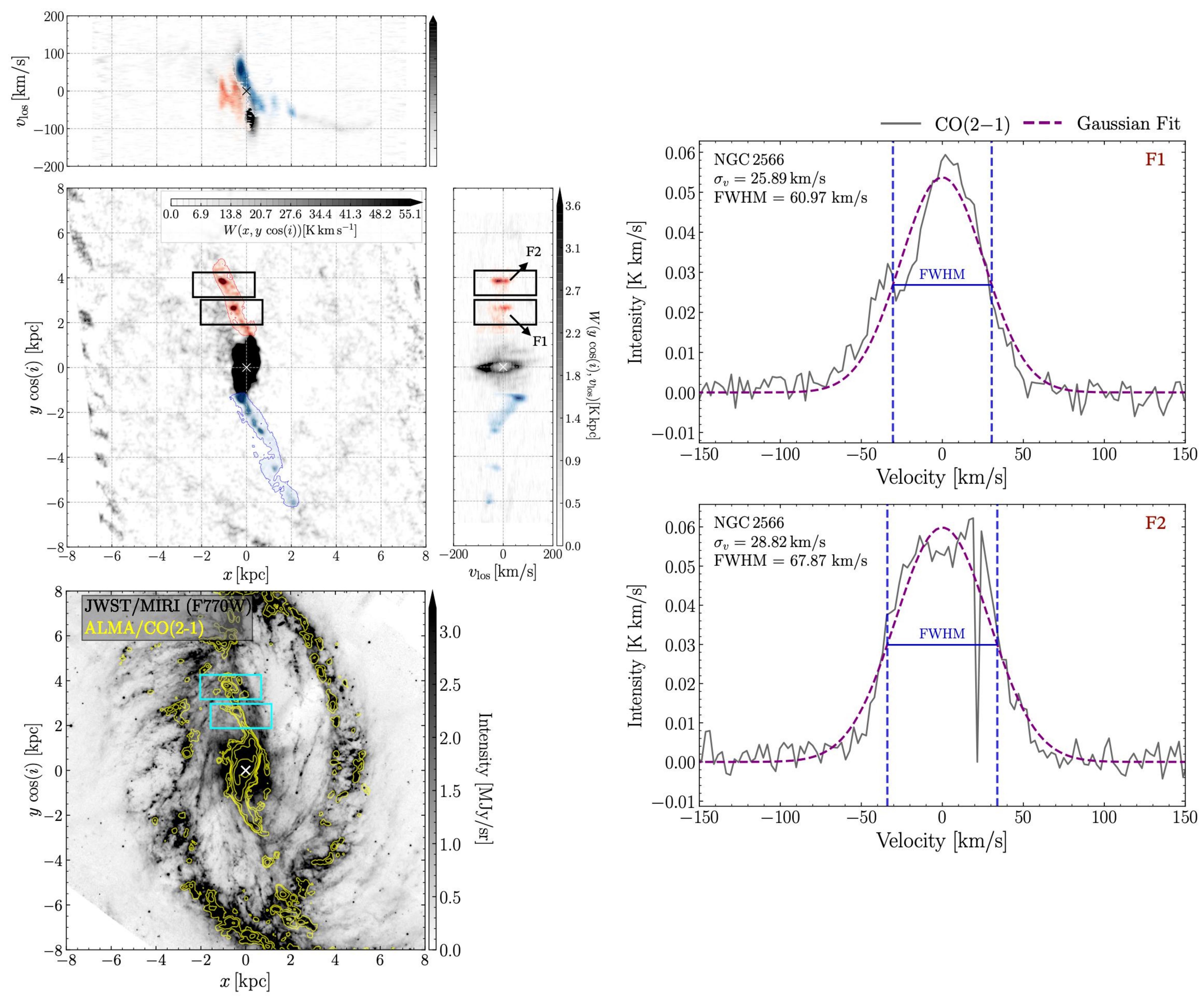}
    \caption{Diagnostic maps, 7.7\,$\rm \mu m$ band observations and spectra within EVF regions for NGC\,2566.}
    \label{fig:NGC2566}
\end{figure*}

\begin{figure*}
    \centering
    \includegraphics[width=1\textwidth]{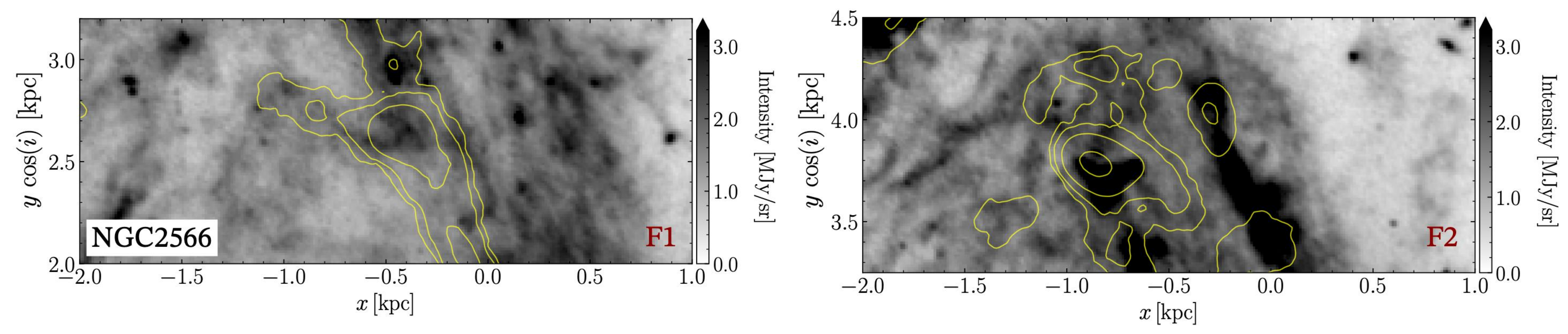}
    \caption{Zoomed-in 7.7\,$\rm \mu m$ band observations of NGC\,2566.}
    \label{fig:NGC2566_b}
\end{figure*}

\renewcommand{\thefigure}{A4\alph{figure}}
\setcounter{figure}{0} 
\begin{figure*}
    \centering
    \includegraphics[width=1\textwidth]{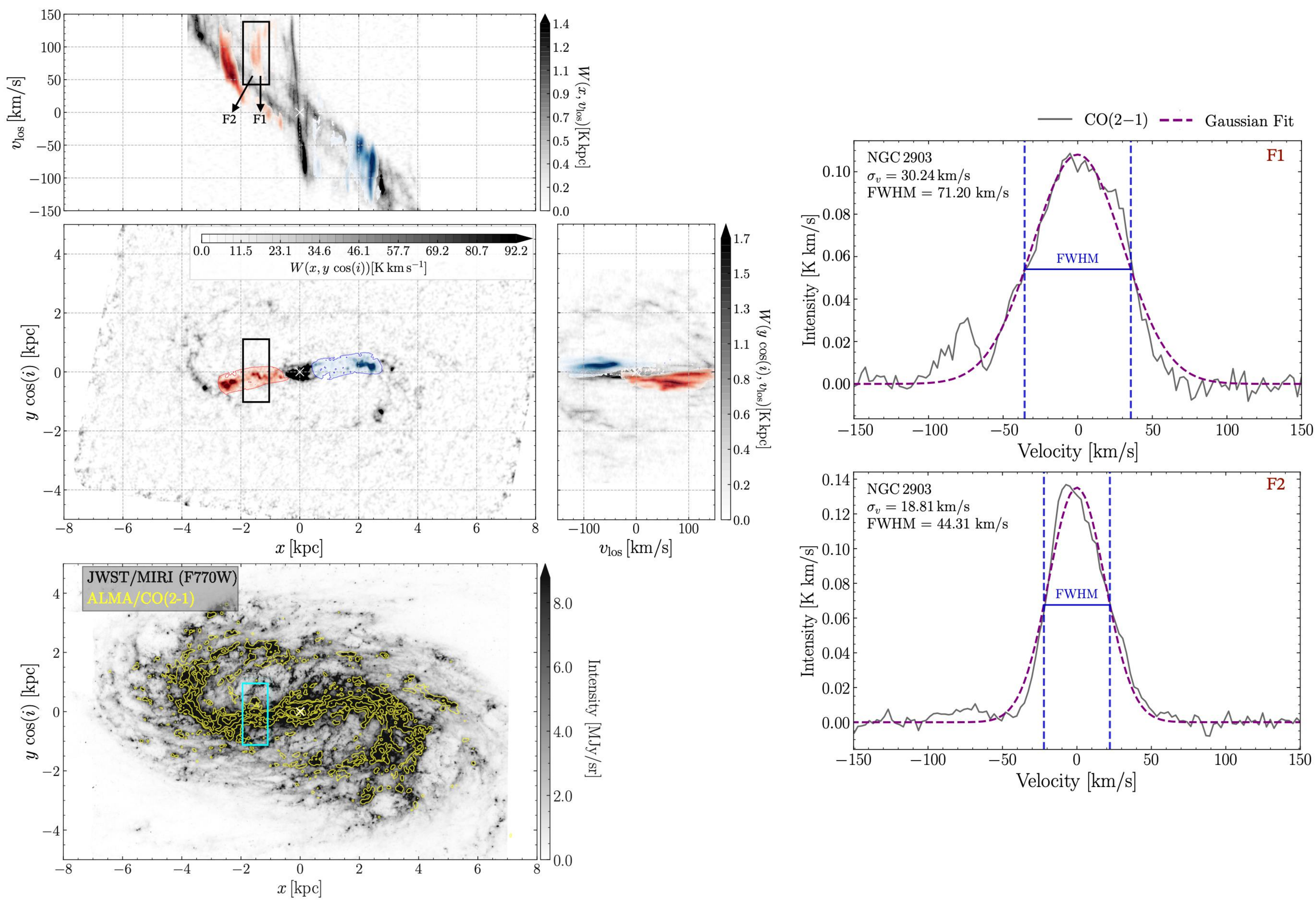}
    \caption{Diagnostic maps, 7.7\,$\rm \mu m$ band observations and spectra within EVF regions for NGC\,2903. The left panels show the same as in Figs.~\ref{fig:NGC1300}, and the right panels show the same as in Fig.~\ref{fig:NGC1300_spectrum}.}
    \label{fig:NGC2903}
\end{figure*}

\begin{figure*}
    \centering
    \includegraphics[width=0.35\textwidth]{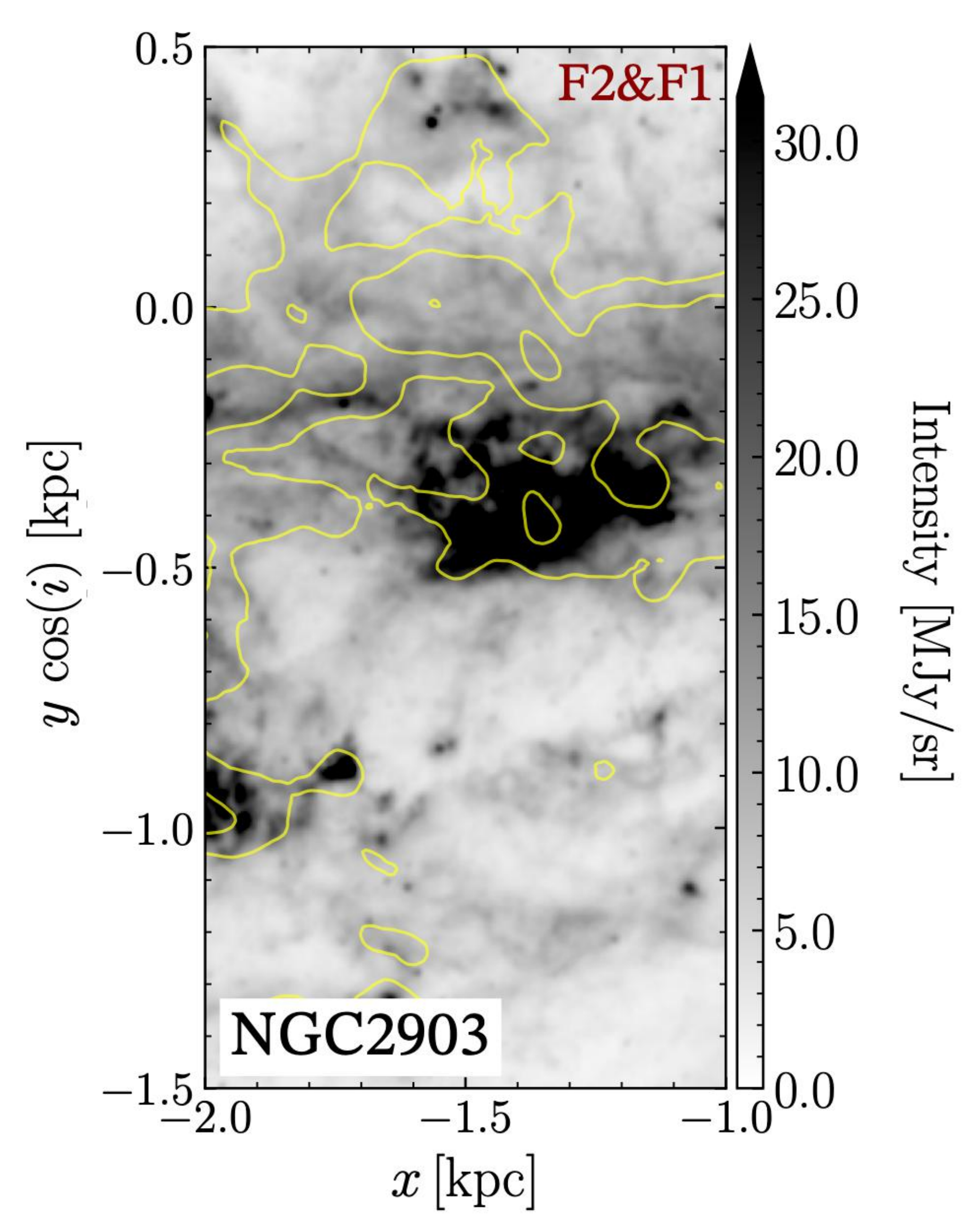}
    \caption{Zoomed-in 7.7\,$\rm \mu m$ band observations of NGC\,2903.}
    \label{fig:NGC2903_b}
\end{figure*}

\renewcommand{\thefigure}{A5\alph{figure}}
\setcounter{figure}{0} 

\begin{figure*}  
\centering

\includegraphics[width=0.9\textwidth]{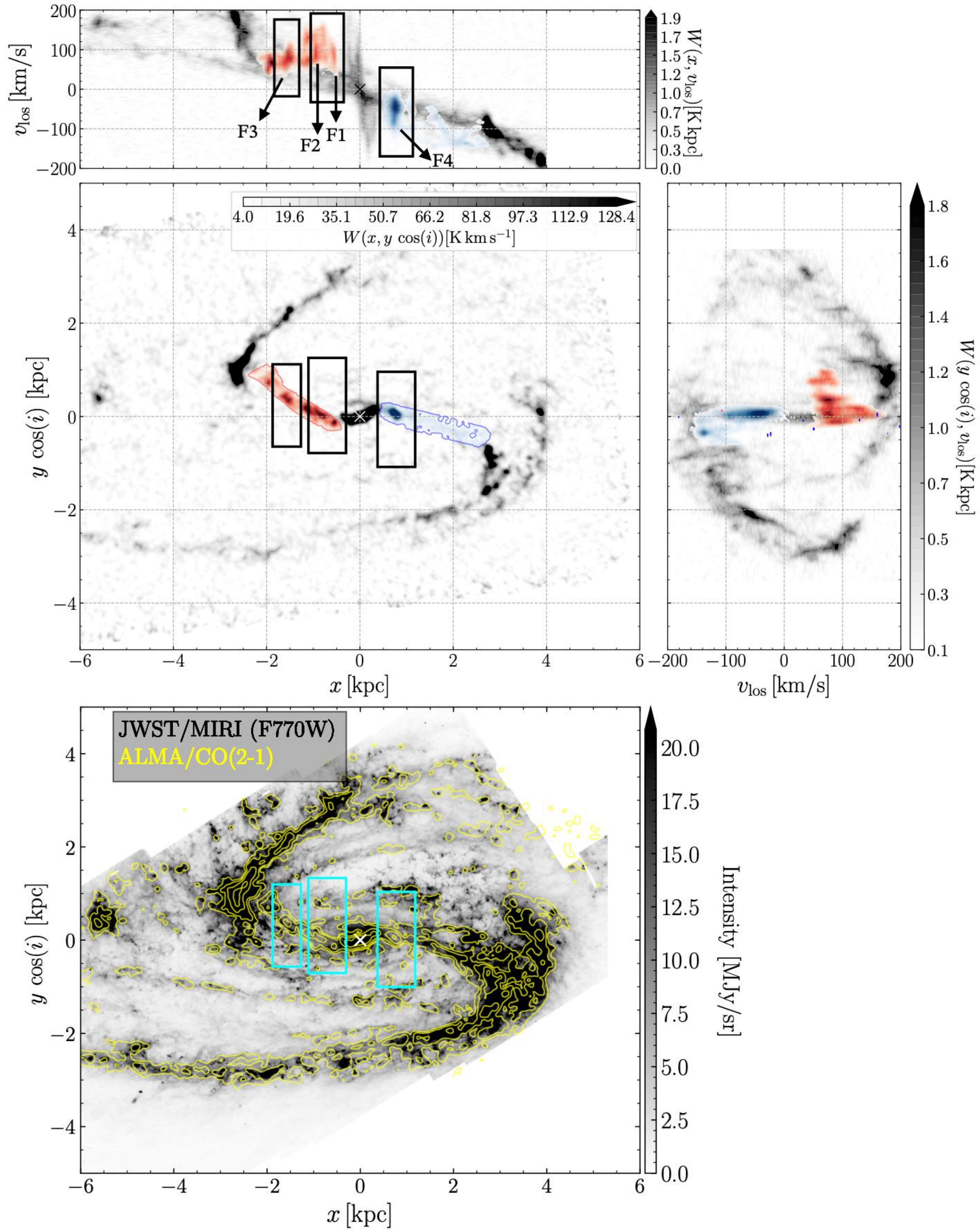}
    \label{fig:NGC3627}
    \caption{Diagnostic maps, 7.7\,$\rm \mu m$ band observations and spectra within EVF regions for NGC\,3627.}
\end{figure*}

\renewcommand{\thefigure}{A5\alph{figure}}
\setcounter{figure}{0} 
\begin{figure*}  
\centering
\includegraphics[width=0.85\textwidth]{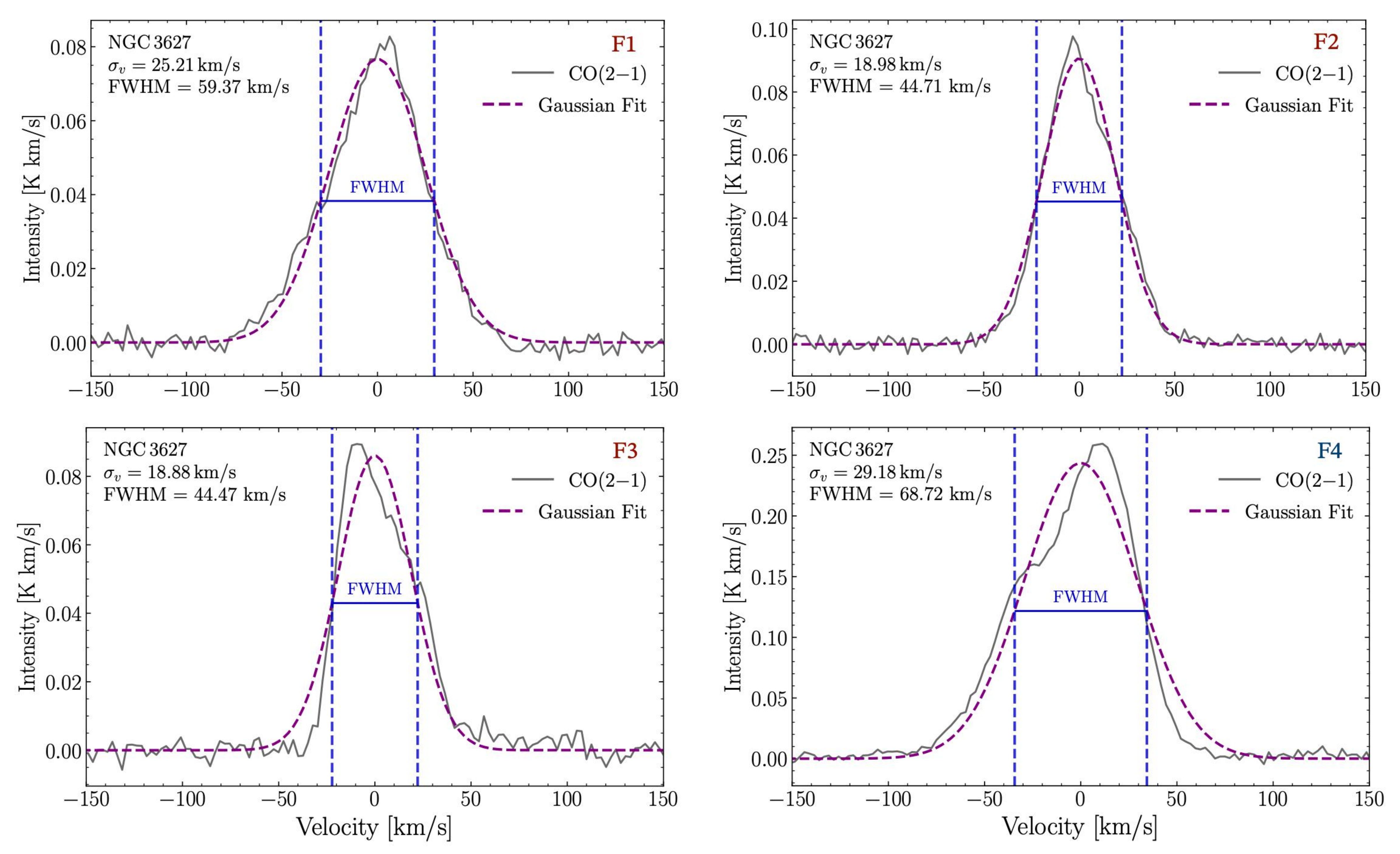}
    \caption{ \textit{Continued}. CO(2--1) spectra within EVF regions for NGC\,3627.}
\end{figure*}

\begin{figure*}
    \centering
    \includegraphics[width=0.9\textwidth]{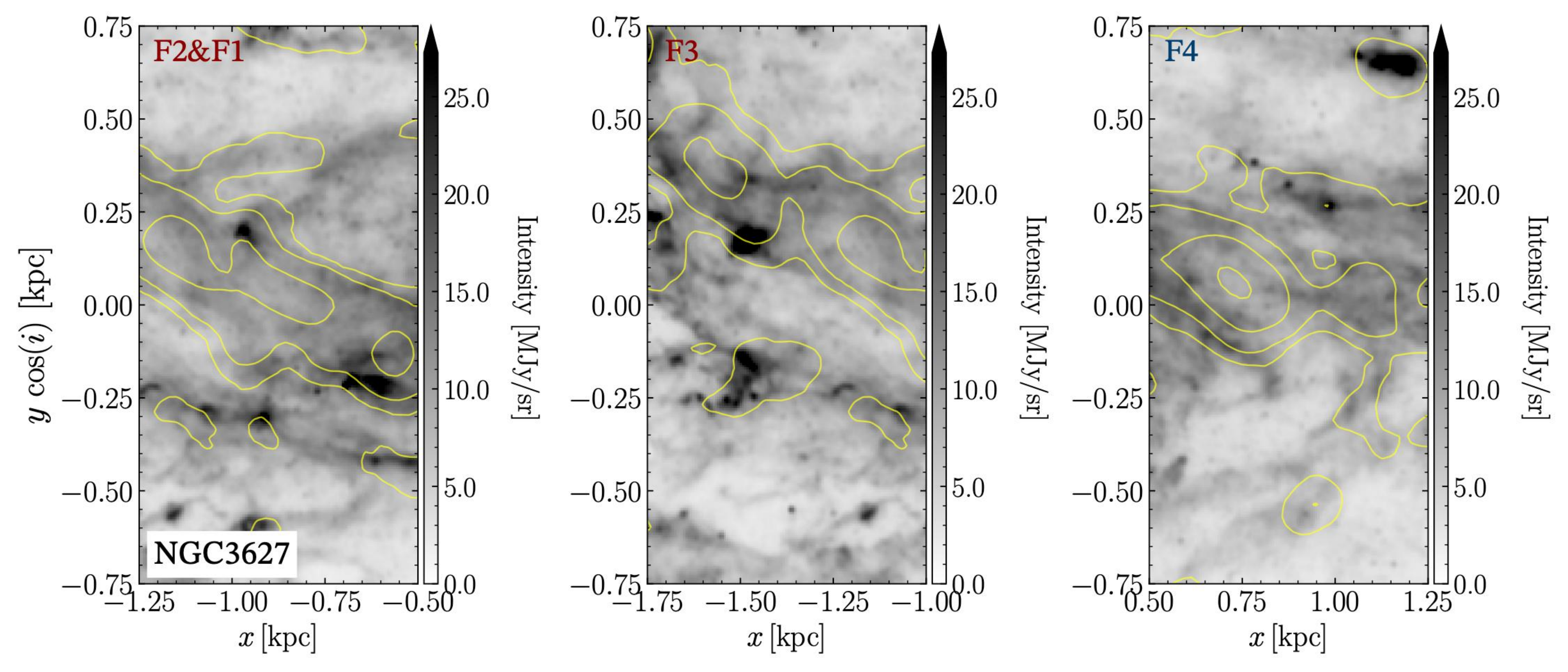}
    \caption{Zoomed-in 7.7\,$\rm \mu m$ band observations of NGC\,3627.}
    \label{fig:NGC3627_b}
\end{figure*}

\renewcommand{\thefigure}{A6\alph{figure}}
\setcounter{figure}{0} 

\begin{figure*}
    \centering
    \includegraphics[width=1\textwidth]{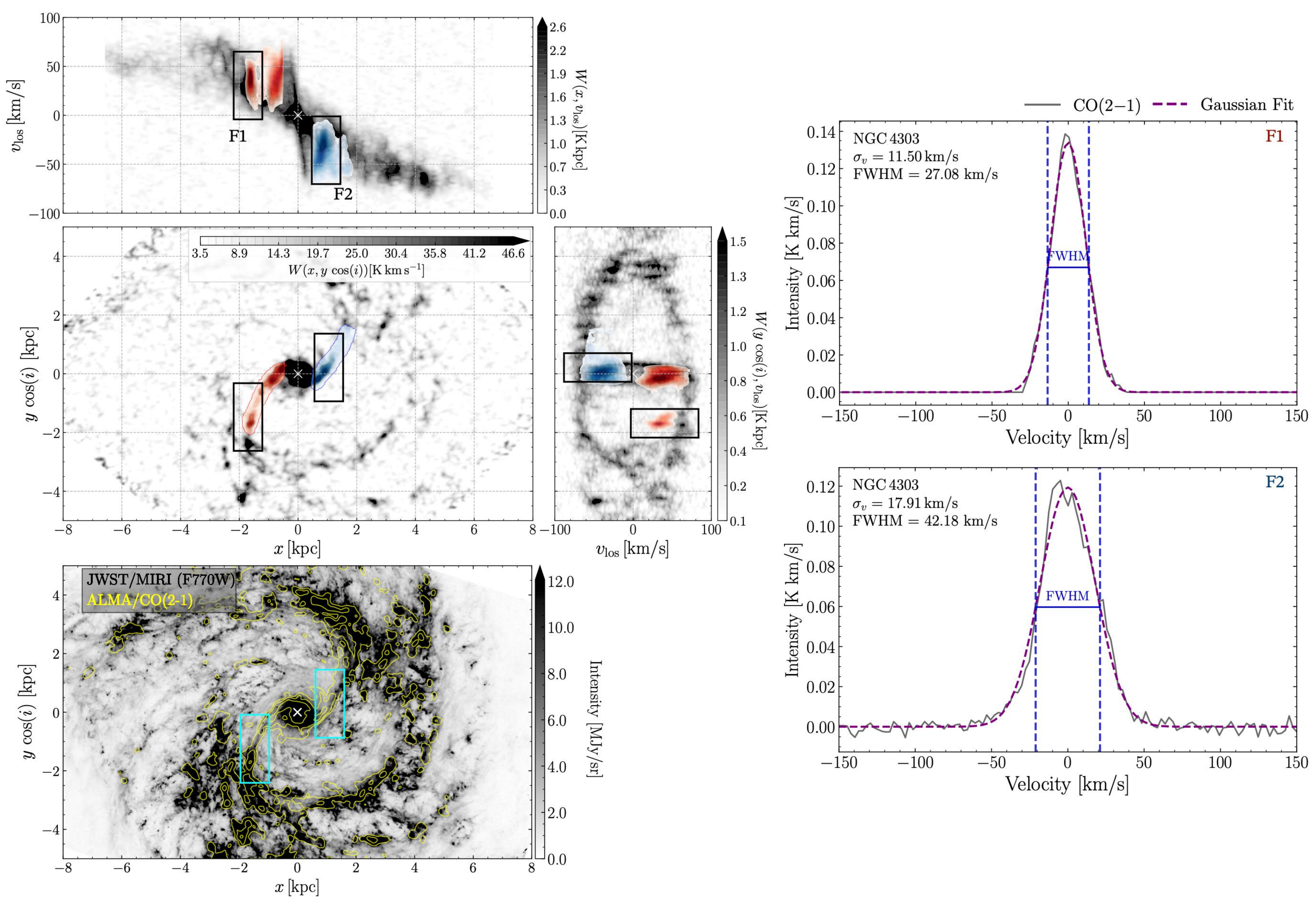}
    \caption{Diagnostic maps, 7.7\,$\rm \mu m$ band observations and spectra within EVF regions for NGC\,4303.}
    \label{fig:NGC4303}
\end{figure*}

\begin{figure*}
    \centering
    \includegraphics[width=0.75\textwidth]{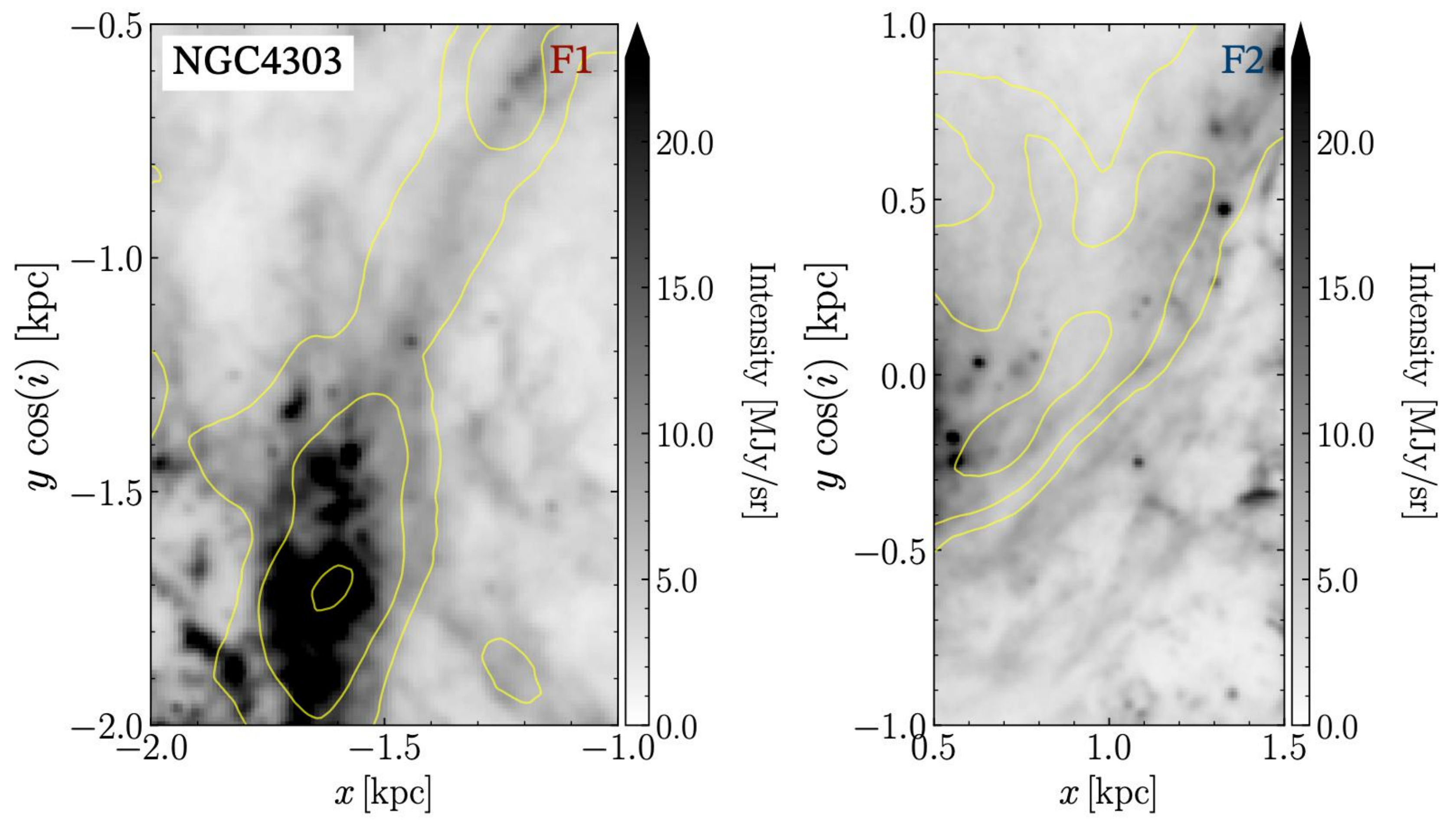}
    \caption{Zoomed-in 7.7\,$\rm \mu m$ band observations of NGC\,4303.}
    \label{fig:NGC4303_b}
\end{figure*}

\renewcommand{\thefigure}{A7\alph{figure}}
\setcounter{figure}{0} 

\begin{figure*}
    \centering
    \includegraphics[width=1\textwidth]{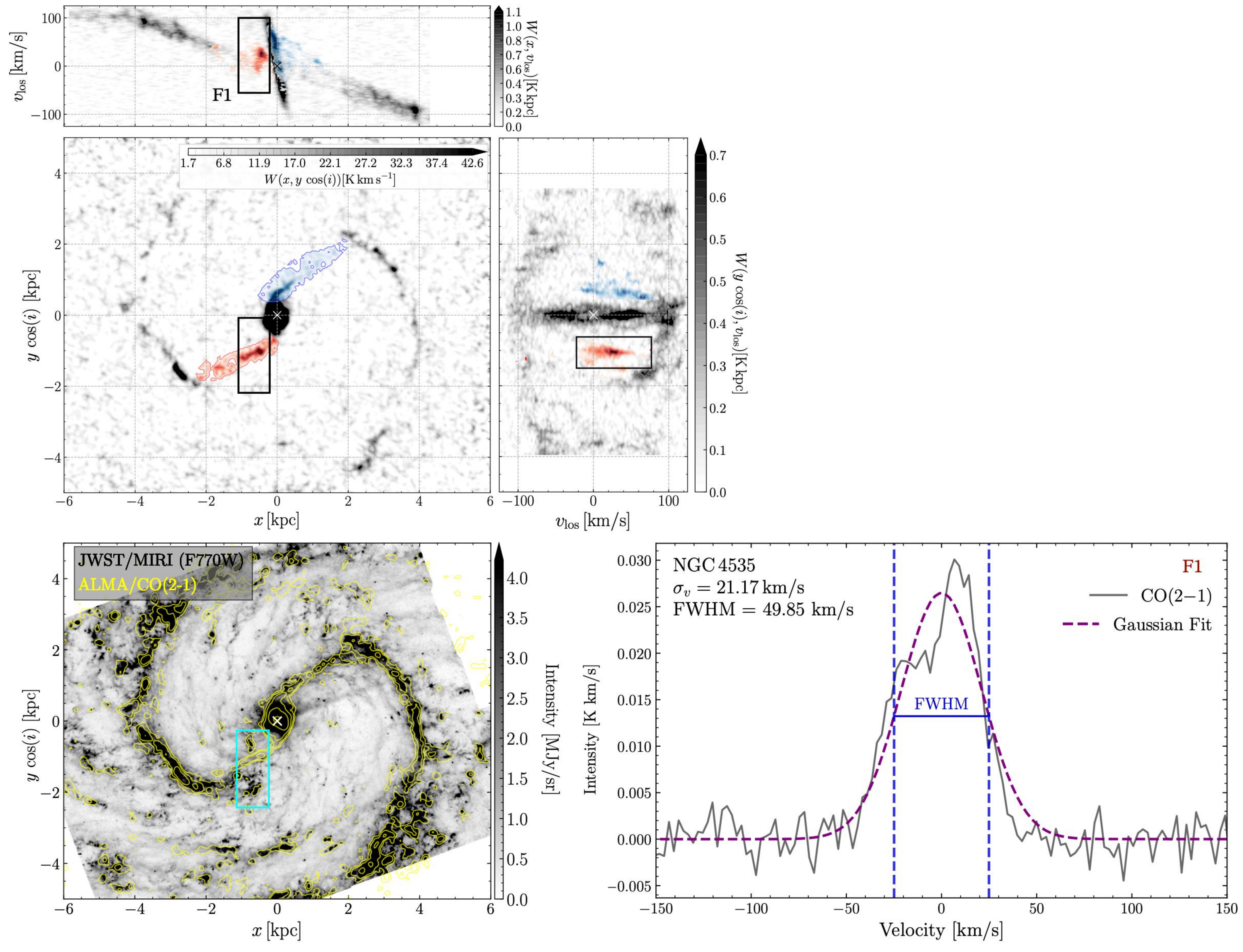}
    \caption{Diagnostic maps, 7.7\,$\rm \mu m$ band observations and spectra within EVF regions for NGC\,4535.}
    \label{fig:NGC4535}
\end{figure*}

\begin{figure*}
    \centering
    \includegraphics[width=0.35\textwidth]{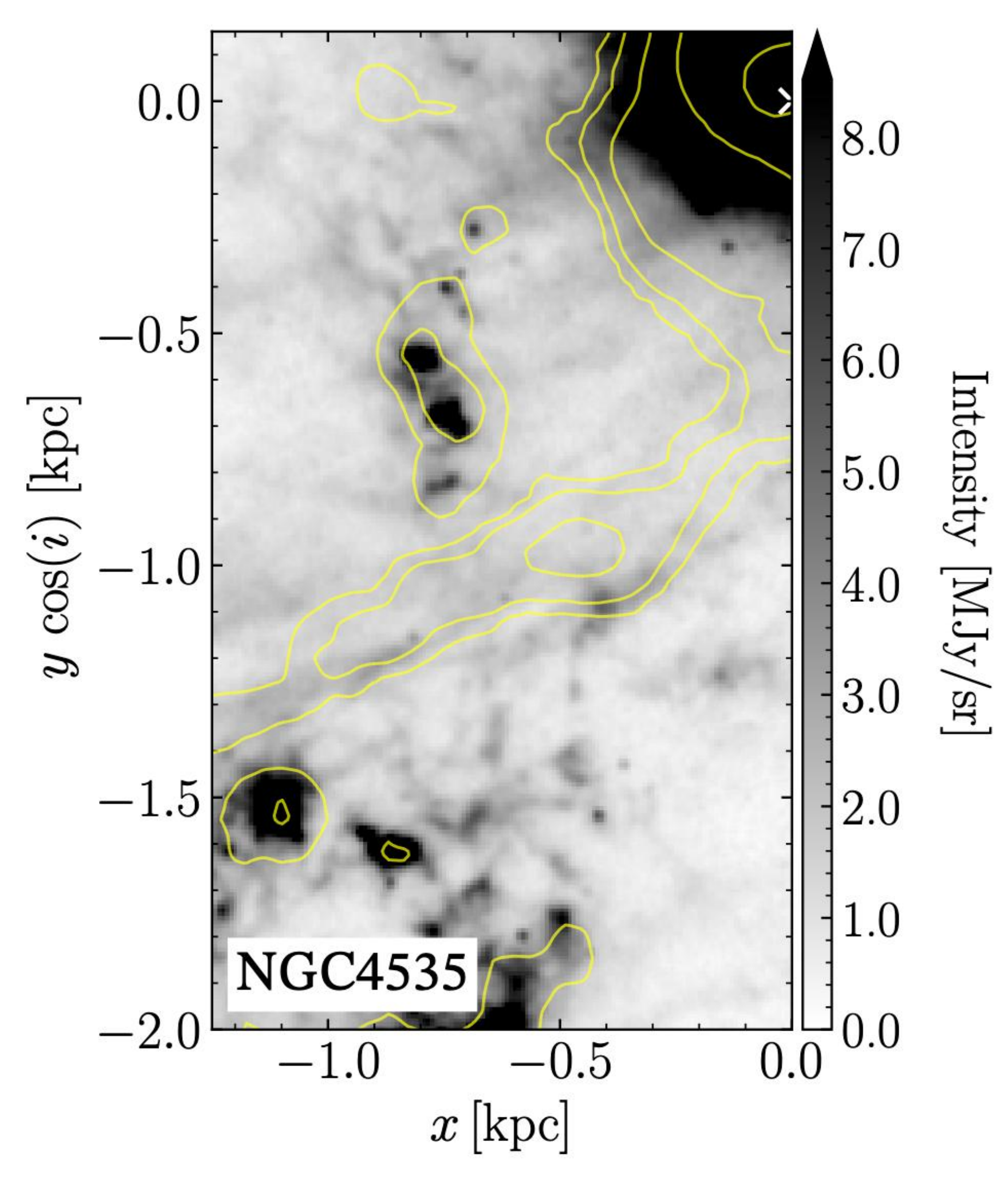}
    \caption{Zoomed-in 7.7\,$\rm \mu m$ band observations of NGC\,4535.}
    \label{fig:NGC4535_b}
\end{figure*}

\renewcommand{\thefigure}{A8\alph{figure}}
\setcounter{figure}{0} 

\begin{figure*}
    \centering
    \includegraphics[width=1\textwidth]{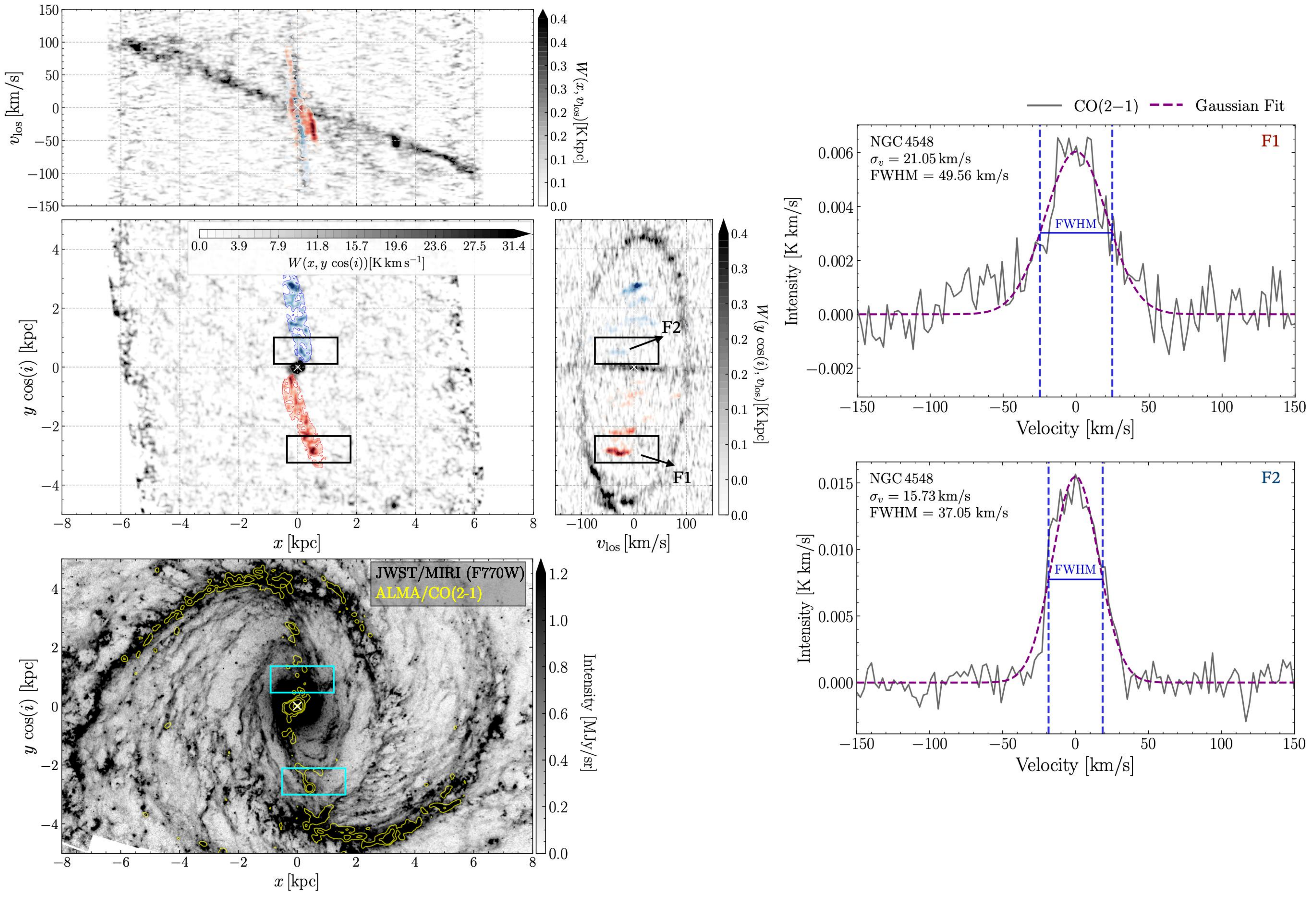}
    \caption{Diagnostic maps, 7.7\,$\rm \mu m$ band observations and spectra within EVF regions for NGC\,4548.}
    \label{fig:NGC4548}
\end{figure*}

\begin{figure*}
    \centering
    \includegraphics[width=0.55\textwidth]{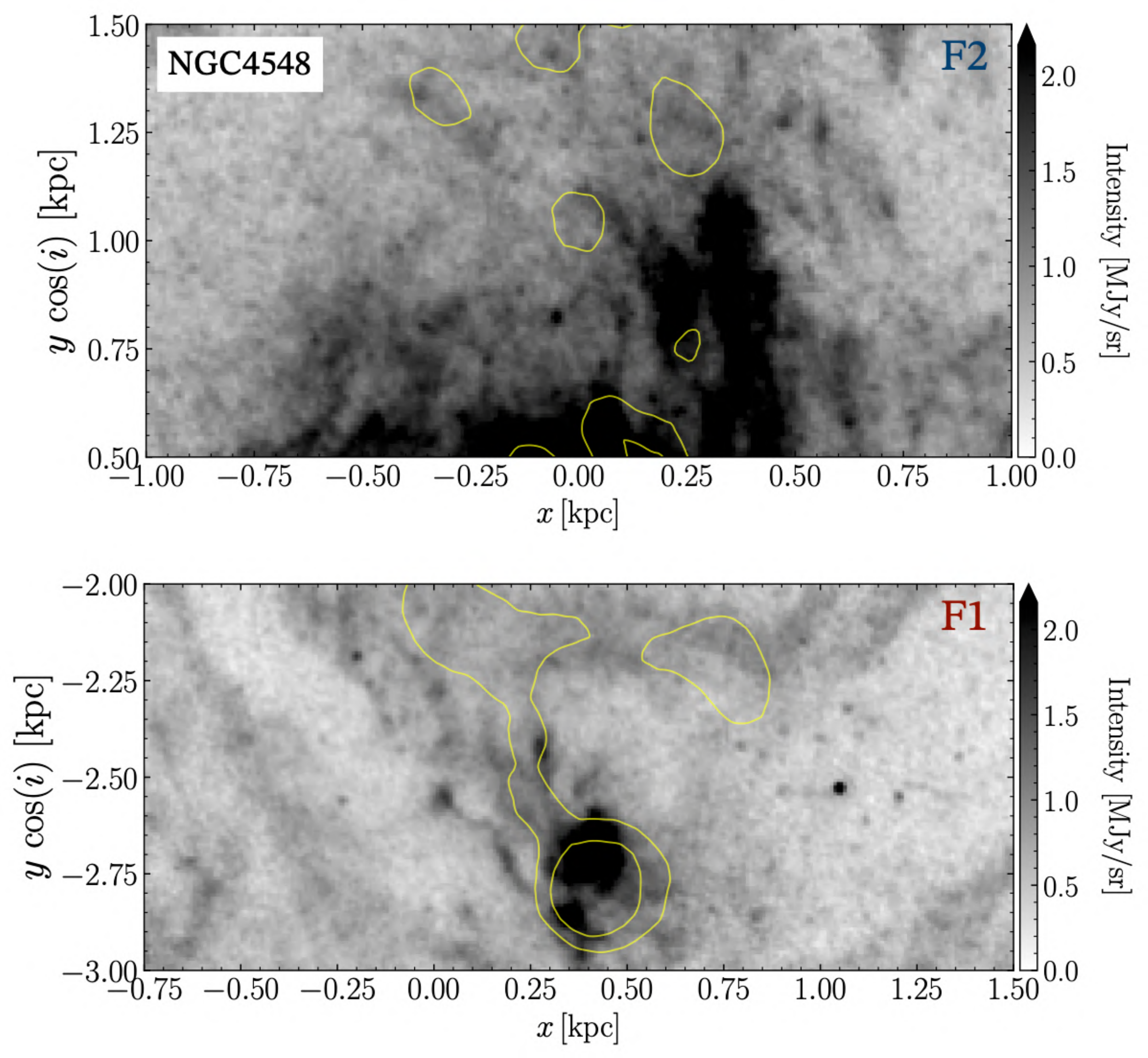}
    \caption{Zoomed-in 7.7\,$\rm \mu m$ band observations of NGC\,4548.}
    \label{fig:NGC4548_b}
\end{figure*}

\renewcommand{\thefigure}{A9\alph{figure}}
\setcounter{figure}{0} 

\begin{figure*}
    \centering
    \includegraphics[width=1\textwidth]{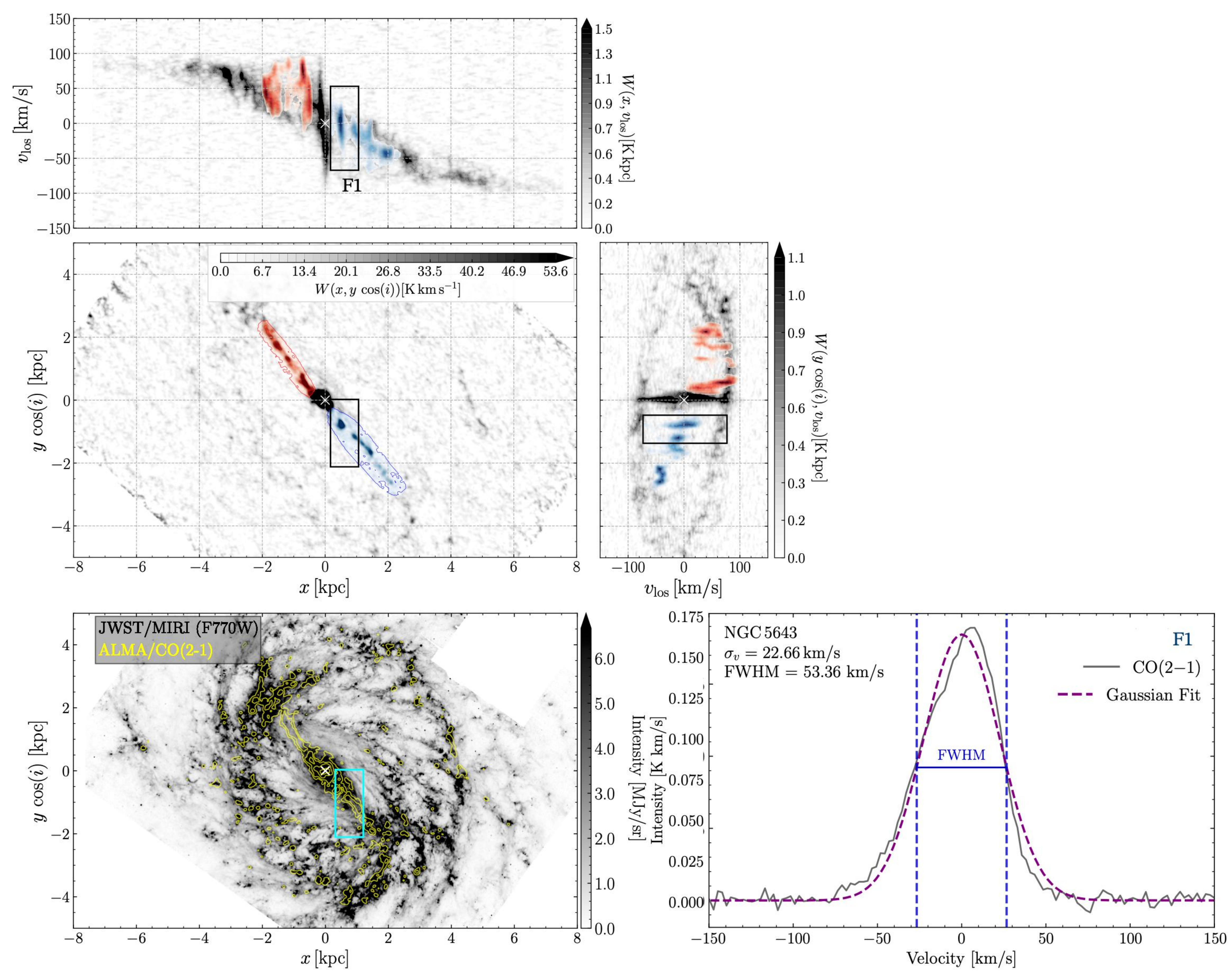}
    \caption{Diagnostic maps, 7.7\,$\rm \mu m$ band observations and spectra within EVF regions for NGC\,5643.}
    \label{fig:NGC5643}
\end{figure*}

\begin{figure*}
    \centering
    \includegraphics[width=0.3\textwidth]{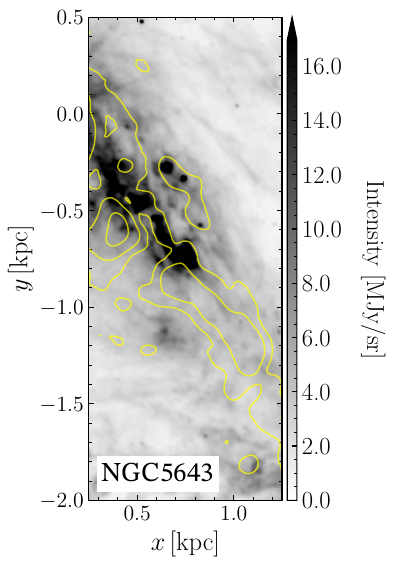}
    \caption{Zoomed-in 7.7\,$\rm \mu m$ band observations of NGC\,5643.}
    \label{fig:NGC5643_b}
\end{figure*}

\end{document}